\documentclass[pra,twocolumn,nofootinbib,floatfix]{revtex4-2}

\usepackage{nameref}
\usepackage{units}
\usepackage{mathtools}
\usepackage{graphicx}
\usepackage{bm}
\usepackage{hyperref}
\usepackage{xcolor}

\makeatletter
\renewcommand\frontmatter@abstractwidth{\dimexpr\textwidth-2in\relax}
\makeatother

\newcommand{\X}[1]{\colorbox{yellow}{S}}

\newcommand{\ket}[1]{\vert {#1} \rangle}
\newcommand{\Er}{E_{\rm r}}

\begin{document}

\title{Commensurate and incommensurate 1D interacting quantum systems}

\author{Andrea~Di~Carli}
\author{Christopher~Parsonage}
\author{Arthur~La~Rooij}
\author{Lennart~Koehn}
\author{Clemens~Ulm}
\author{Callum~W~Duncan}
\author{Andrew~J~Daley}
\author{Elmar~Haller}
\author{Stefan~Kuhr}\email{stefan.kuhr@strath.ac.uk}

\date{\today}

\affiliation{\vspace{0.3cm}Department of Physics, SUPA, University of Strathclyde, Glasgow G4 0NG, United Kingdom}


\begin{abstract}
\vspace{0.2cm}
Single-atom imaging resolution of many-body quantum systems in optical lattices is routinely achieved with quantum-gas microscopes. Key to their great versatility as  quantum simulators is the ability to use engineered light potentials at the microscopic level. Here, we employ dynamically varying microscopic light potentials in  a quantum-gas microscope to  study commensurate and incommensurate 1D systems of interacting bosonic Rb atoms.
Such incommensurate systems are analogous to doped insulating states that exhibit atom transport and compressibility.
Initially, a commensurate system with unit filling and fixed atom number is prepared  between two potential barriers. We deterministically create  an incommensurate system by dynamically changing the position of the barriers such that the number of available lattice sites is reduced while retaining the atom number.  Our systems are characterised by  measuring the distribution of particles and holes as a function of the lattice filling, and interaction strength, and we probe the particle mobility by applying a bias potential. Our work provides the foundation for preparation of low-entropy states with controlled filling in optical-lattice experiments.
\end{abstract}


\date{\today}

\maketitle

\subsection*{\label{sec:Intro}Introduction}
Quantum-gas microscopes \cite{Bakr2010,Sherson2010,Gross2021} offer a unique tool for quantum simulation
of  many-body quantum systems \cite{Lewenstein2007,Bloch2008,Gross2017,Schafer2020} in optical lattices with single-atom imaging resolution and control.
The addition of tailored static light potentials has made it possible to create box-like traps and cancel the harmonic confinement, enabling the study of homogeneous systems \cite{Kaufman2016,Lukin2019} and topological and magnetic phases in restricted dimensions \cite{Mazurenko2017,Sompet2022,Hirthe2023}.
It is experimentally challenging to control the number of particles \cite{Wenz2013} and the filling, which can dramatically change the properties of the quantum system. The addition or removal of a particle is analogous to doping in semiconductors, and it is relevant to the physics of doped antiferromagnet high-T$_{\rm c}$ superconductors \cite{Lee2006}. Recent experimental studies using quantum-gas microscopes have shed light on the role of doping in Fermi-Hubbard systems \cite{greiner2023}, by observing bulk transport properties \cite{Anderson2019,Brown2019,Nichols2019},  string patterns \cite{Chiu2019},  incommensurate magnetism \cite{Salomon2019}, magnetic polarons \cite{Koepsell2019,Koepsell2021,Ji2021}, and hole pairing via magnetic coupling \cite{Hirthe2023}.

The lattice filling is equally  relevant for bosonic many-body quantum systems. In the case of a commensurate particle number, i.e., an integer filling fraction,  a homogeneous system can attain a Mott-insulating phase \cite{Fisher1989,Jaksch1998}, while for systems with incommensurate fillings
interesting quantum phases have been predicted, such as supersolid and crystalline phases \cite{Lazarides2011,Buchler2011} and the Bose-glass phase in the presence of a disordered potential \cite{Fisher1989,Damski2003,Fallani2007,Cai2010}, or  defect-induced superfluidity \cite{Astrakharchik2017}. Dynamic control over the shape of the light potential using a digital micromirror device (DMD) has recently been used to stir quantum gases to generate vortices \cite{Kwon2021,Pace2022,Reeves2022} and to switch between two optical potentials \cite{Ji2021,Hirthe2023}.

In this work, we use a DMD to create dynamic light potentials at a microscopic level to control the commensurability of bosonic quantum systems. We
initially prepare commensurate 1D bosonic quantum systems at unit filling between repulsive potential barriers, before the confining potential is dynamically changed to reduce the number of available lattice sites while retaining the atom number. With this incommensurate filling, the system is no longer a Mott insulator, which we probe by studying the mobility of particles when subjected to a bias potential. The incommensurate systems with delocalised atoms on a localised background also feature nontrivial
site occupation probabilities in
the ground state \cite{Brouzos2010}. Our technique to use  a dynamic light potential to deterministically prepare a low-entropy incommensurate quantum state can also be applied to study many-body quantum systems in different confining potentials or lattice geometries \cite{Yamamoto2020,Yang2021}.

\begin{figure}[!t]
\centering
\includegraphics[width=1\columnwidth]{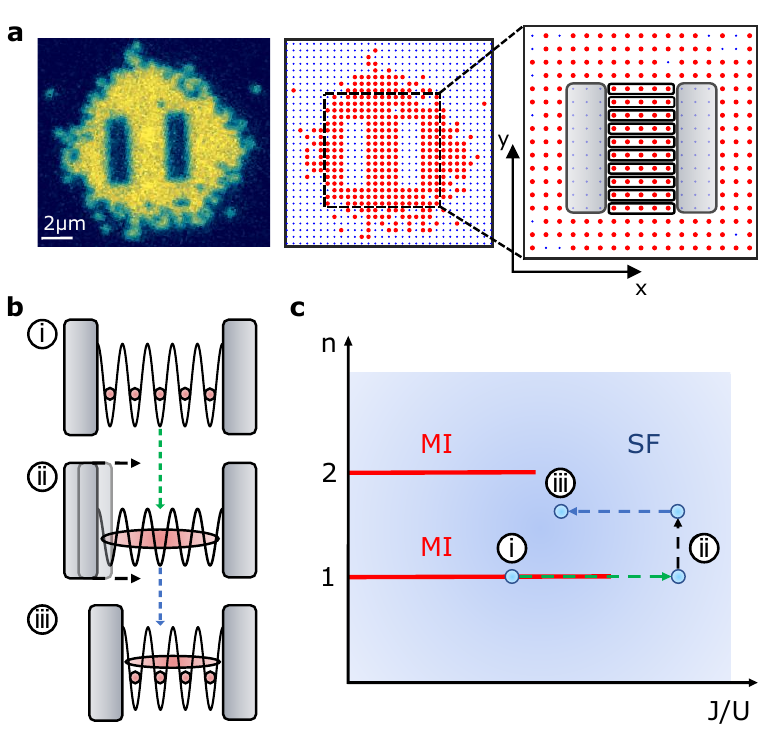}
\caption{\label{fig:concept}\textbf{1D system preparation, experimental scenario and phase diagram.} \textbf{a}, Left: fluorescence image of a Mott insulator of $^{87}$Rb atoms in the presence of two repulsive potential barriers, visible as hollow rectangles in the centre,  Middle:  corresponding atom distribution. Right: Magnification of the central region, highlighting the individual 1D systems with five atoms and the location of the repulsive potential (grey shaded areas).
\textbf{b}, Sketch of the procedure to generate an incommensurate (doped) 1D quantum system: (i) Initial preparation of a Mott insulating state, (ii) transition to the superfluid regime, and  reduction of the number of lattice sites  by moving the potential barrier,  (iii) transition into the strongly interacting regime. \textbf{c}, Illustration of the phase diagram for the 1D Bose-Hubbard model for finite particle number indicating the path followed through stages (i)-(iii). }
\end{figure}

\subsection*{Preparation of (in)commensurate systems}
To prepare and detect our 1D commensurate and incommensurate quantum systems with up to 6 atoms, we employ a quantum-gas microscope which allows for single-atom-resolved detection of  bosonic $^{87}$Rb atoms, using  a setup similar to earlier studies \cite{Sherson2010,Fukuhara2013a}. Initially, we create a degenerate 2D quantum gas of around $400$ atoms in a single antinode of a vertical optical lattice in the $z$-direction  overlapped with  two horizontal optical lattice beams (wavelength $\lambda=1064$\,nm) in the $x$- and $y$-directions (Methods). Using a DMD and light at 666\,nm wavelength, we produce two repulsive  potential barriers of rectangular shape covering $3 \times 10$ lattice sites each  (Fig.\,\ref{fig:concept}a) that are projected onto the optical lattice by a high-resolution microscope objective.
We create ten independent commensurate one-dimensional systems with unit filling by initially preparing a  2D Mott-insulating state. This is done by changing the lattice potential of both horizontal beams from $0$ to $V_x=V_y=50(2)\,E_{\text{r}}$ within 500\,ms. Here,  $E_{\rm r}=\hbar^2/2m\lambda^2$ is the recoil energy, with $m$ being the atomic mass of $\prescript{87}{}{\textrm{Rb}}$. During the quantum phase transition from a superfluid to the Mott insulator, the open geometry in the $y$-direction of the repulsive  potential allows for the redistribution of residual entropy towards the outer regions thus enhancing the preparation fidelity in the centre.

The   experimental procedure  is illustrated in Fig.\,\ref{fig:concept}b together with a phase diagram in Fig.~\ref{fig:concept}c.
Initially, our commensurate 1D system in a Mott-insulating state (Fig.~\ref{fig:concept}b, panel i) is brought into the superfluid regime (Fig.~\ref{fig:concept}b, panel ii). The position of the repulsive potential barrier is then moved to reduce the number of available lattice sites while retaining the atom number (Fig.~\ref{fig:concept}b, panel iii). As a result, when the 1D system is brought back into the strongly interacting regime, it can no longer form a Mott insulator with unit filling, as shown in the phase diagram (Fig.~\ref{fig:concept}c).

\begin{figure*}
\includegraphics[width=0.80\textwidth]{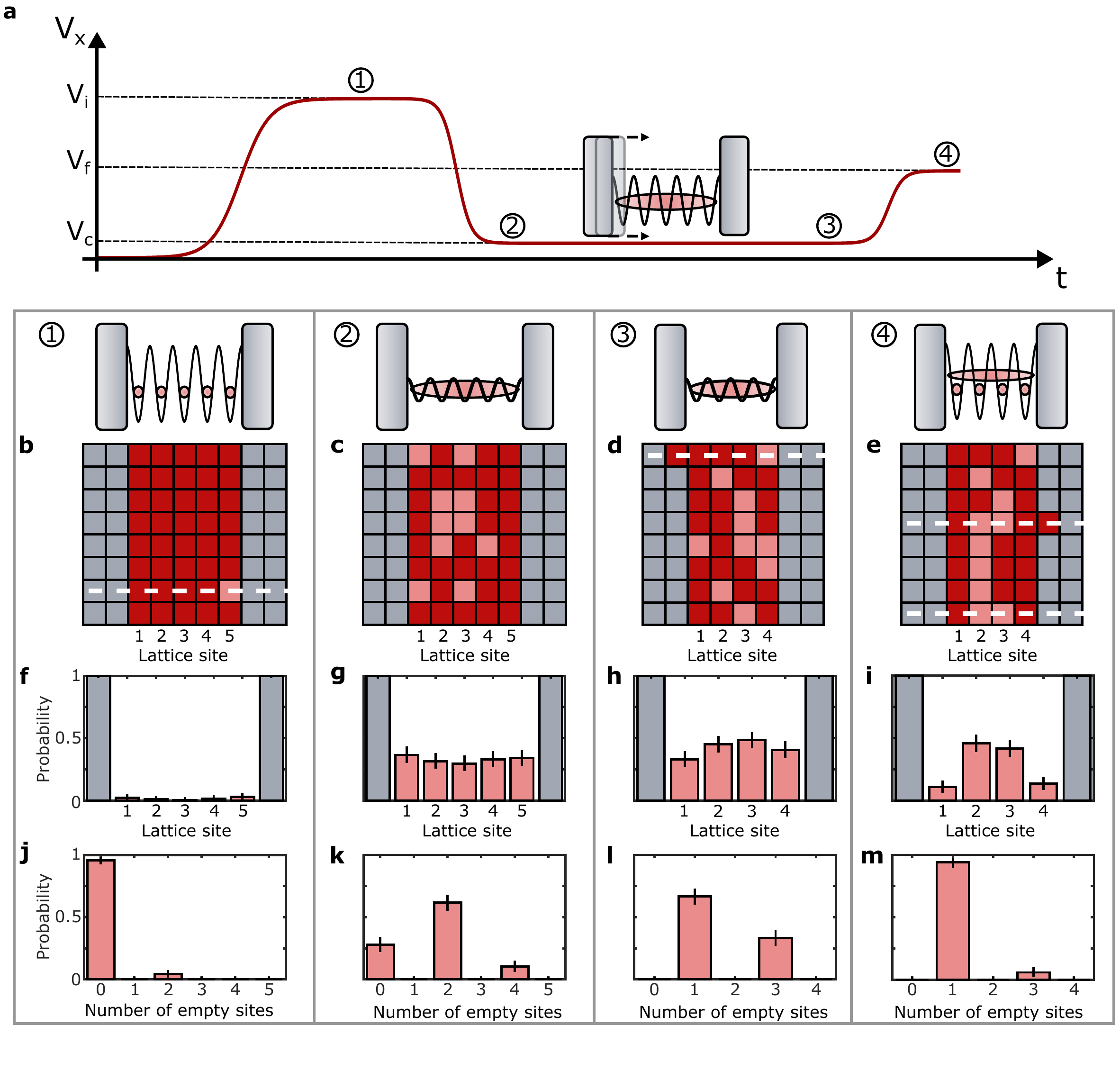}
\vspace{-0.5cm}
\caption{\label{fig:DopingSequence}\textbf{Experimental procedure for doping a Mott insulator.} \textbf{a}, Time-dependent variation of the $x$-lattice potential, $V_x$. Numbers in circles indicate the stages at which measurements are performed:  (1) after preparation of a commensurate system in a  Mott-insulating state at $V_i=50(2)\,\Er$; (2) after preparation of a superfluid state in a shallow lattice, $V_c=2.8(4)\,\Er$; (3) after creating an incommensurate system by dynamically compressing the superfluid, (4) after increasing the lattice depth to reach the strongly interacting regime again, $V_f=16(1)\,E_{\rm r}$. \textbf{b-e}, Reconstructed lattice occupation of one experimental realisation, showing the repulsive potential (grey), atoms (dark red) and observed empty sites  (light red) that result from both holes and doublons. White dashed lines indicate rows excluded from the statistics by post-selection. \textbf{f-i}, Observed probabilities of detecting an empty site. \textbf{j-m}, Probability vs number of empty sites for the same system.  Each histogram is obtained by averaging over 260-380 independent 1D  systems, and all error bars are the 95\% Clopper-Pearson confidence intervals.\\
}
\end{figure*}

To characterise the commensurate and incommensurate 1D systems at each stage of the experimental sequence (Fig.\,\ref{fig:DopingSequence}a),
we record the parity of the atom number (Fig.\,\ref{fig:DopingSequence}b-e) on each lattice site (Methods), as due to light-assisted collisions during the fluorescence imaging we measure the  atom number modulus two \cite{Sherson2010}.
From this, we calculate the probability of finding empty sites  as a function of the  lattice site (Fig.\,\ref{fig:DopingSequence}f-i) and a histogram of the number of empty sites  (Fig.\,\ref{fig:DopingSequence}j-m) per 1D system.  We post-select the datasets by excluding 1D systems (white dashed lines Fig.\,\ref{fig:DopingSequence}b-e) in which the wrong parity is measured   or in which an atom is detected at the position of the  potential barrier (Supplementary Information). After this post-selection we retain  on average $70$\% of the 1D systems, creating effectively a low-temperature subset of the measured datasets \cite{Fukuhara2013a}.

%
%

We initially measure  the preparation fidelity of five atoms on five lattice sites in the strongly interacting regime. In this scenario  with commensurate filling, each atom is localised on a single lattice site. We measure 96(2)\% of the systems with the expected atom number (Fig.~\ref{fig:DopingSequence}j), and in 4(2)\% of the cases we find two empty sites equally distributed across the system (Fig.~\ref{fig:DopingSequence}f). We attribute these to our non-zero initial temperature and to excitations arising from technical noise.
To enter the superfluid regime, the $x$-lattice potential is decreased from  $V_i=50(2)\,\Er$ to $V_c=2.8(4)\,\Er$ within 150\,ms, thereby increasing $J/U$. Here, $J$ is the tunnelling rate, and $U$ is the onsite interaction in the Bose-Hubbard model (Methods). We keep the $y$-lattice at $V_y=50(2)\Er$ to prevent tunneling between the 1D systems. The atoms within the superfluid 1D system become delocalised and due to the atom number fluctuations, we observe an increased number of  empty sites  which have a uniform spatial  distribution  (Fig.~\ref{fig:DopingSequence}g). We now move the position of one of the potential barriers in 18 discrete steps (Methods), within \unit[200]{ms}. The system size is reduced to four sites while retaining the  five initial atoms, creating a doped system with incommensurate filling. As a consequence, we observe an odd number of empty sites  (Fig.~\ref{fig:DopingSequence}l).  Then, the $x$-lattice potential is increased to $V_f=\unit[16(1)]{E_r}$ within 200\,ms to bring the incommensurate systems back into the strongly interacting regime, leading to the suppression of holes (Fig.~\ref{fig:DopingSequence}m).  The  distribution of empty sites, which now correspond to sites occupied by two atoms, shows a higher probability on the central two sites (Fig.~\ref{fig:DopingSequence}i). The occupation of the central sites is energetically favourable due to the boundary, as predicted by our simulations of the single-band Bose-Hubbard model (Methods).

\subsection*{Strongly and weakly interacting systems}
%
%
\begin{figure*}[]
\includegraphics[width=0.75\textwidth]{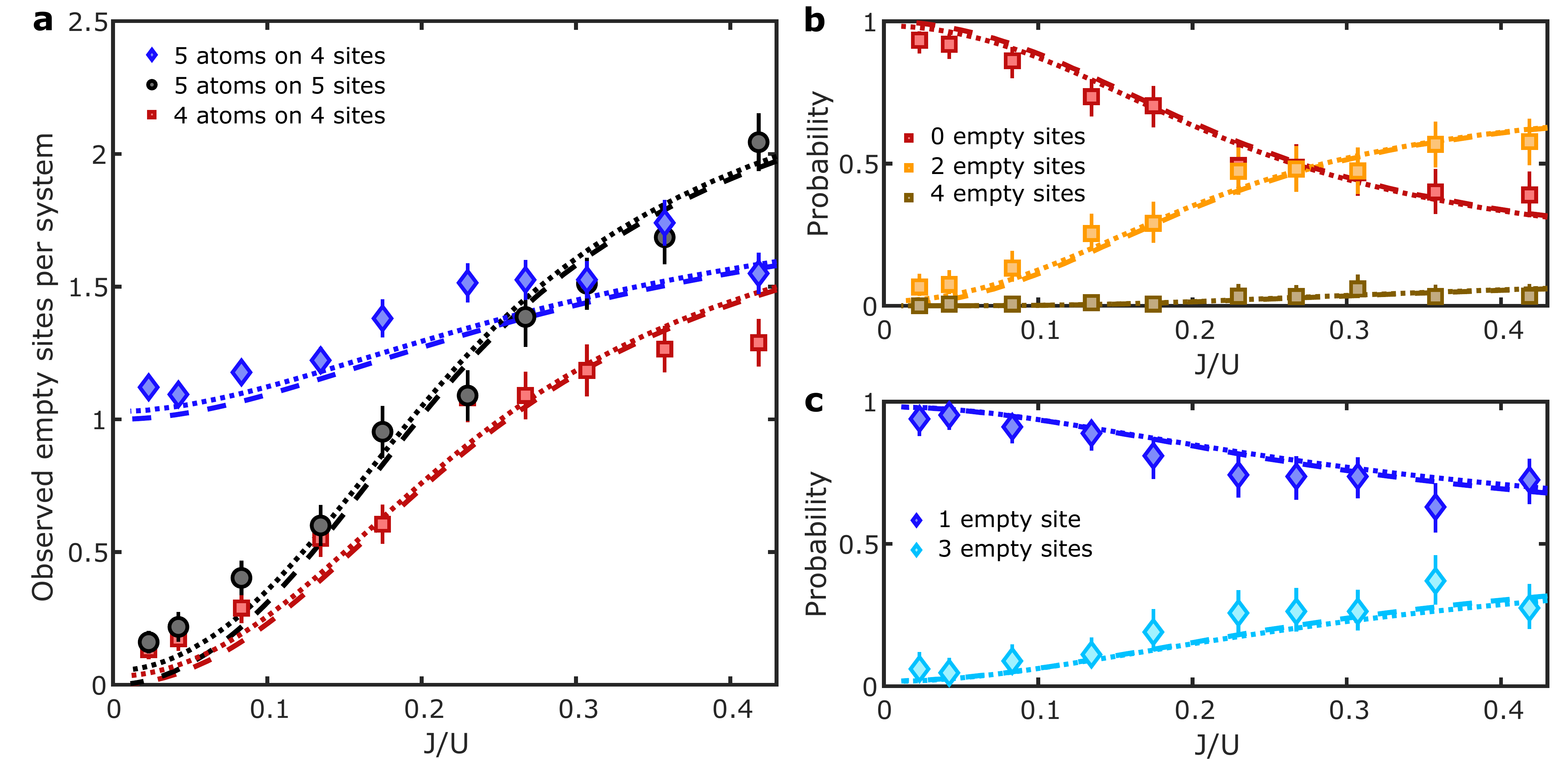}
\caption{\label{fig:JovU}
\textbf{Strongly and  weakly interacting commensurate and incommensurate 1D systems.} \textbf{a},  Number of empty sites per 1D system vs $J/U$ for 5 atoms on 5 lattice sites (black circles), 4 atoms on 4  sites (red squares), 5 atoms on 4 sites (blue diamonds). Observed empty sites  result from both holes and doublons.  Error bars show the standard error.
 Our numerical simulations show the number of empty sites  for $T=0$ (dashed lines) and for $T=0.15\,U$ (dotted lines).
 \textbf{b}, Probabilities of zero (red), two (orange) and four empty sites (brown) per 1D system vs $J/U$ in a commensurate system with 4 atoms on 4 sites. \textbf{c}, same for an incommensurate system  with 5 atoms on 4 sites, showing the probabilities of finding one (blue) and three  (cyan) empty sites. Each data point is obtained by averaging over $110-190$ independent 1D quantum systems, and error bars in \textbf{b} and \textbf{c} are the Clopper–Pearson 95\% confidence intervals. }
\end{figure*}

We study the difference between an incommensurate and commensurate 1D system when transitioning from the weakly to the strongly interacting regime. Specifically, we prepare an incommensurate  system with five atoms on  four lattice sites, and compare it to one with commensurate fillings of five and four atoms on five and four sites, respectively (Fig.\,\ref{fig:JovU}a). We use the same experimental procedure to prepare the commensurate and incommensurate systems, the only difference being that  repulsive barriers are not moved for the commensurate ones. The number of observed empty sites is compared to our numerical simulations (Fig.~\ref{fig:JovU}), taking into account the time-varying potential during the entire experimental procedure (Methods), for both $T=0$ and $T=0.15\,U$. The latter is the measured temperature of the initial 2D Mott insulator, which is an upper bound because the effective temperature for the 1D systems is lower as a result of the reduced entropy in the centre region (Fig.~\ref{fig:concept}c) and the post-selection.

In the strongly interacting regime, $J/U \ll 1$, we observe on average less than 0.2 empty sites in the commensurate system, as it attains a Mott-insulating state. In contrast, in the incommensurate system, we observe close to one empty site (Fig.\,\ref{fig:JovU}a), due to the appearance of a doubly occupied site resulting from one delocalised atom on a localised background. As we increase $J/U$ to enter the weakly interacting or superfluid regime, the number of observed empty sites increases in all three cases, in good agreement with the numerical simulation (Methods).
Using the same data sets, we evaluated the probabilities of detecting empty sites in each 1D system (Fig.~\ref{fig:JovU}b and c). As $J/U$ is increased, we observe that for the commensurate system with 4 atoms on 4 sites, the probability of observing zero empty sites decreases below 0.5 while the occurrence of two empty sites increases accordingly. This is well captured by the numerical simulation that take into account the intensity ramps used to change the lattice depths. In the case of incommensurate filling, the increase in the observed number of empty sites is less pronounced, as the expected number of empty sites per 1D system in the superfluid is only $\approx 1.6$ at zero temperature.
In the strongly interacting regime, we observe more empty sites in the commensurate system than predicted by  the finite temperature simulations (Fig.\,\ref{fig:JovU}a), which is attributed to  loss of two atoms and particle-hole pair excitations \cite{Endres2011} due to heating from intensity noise on the trapping lasers. In the incommensurate system, these effects are more prominent due to the non-gapped excitation spectrum. Assuming that in 30\% of 1D systems we lose one atom, a loss of two atoms is expected 10\% of the time, and these two-atom losses are not accounted for in the post-selection as the parity is conserved.

\subsection*{Atom number,  variance and site occupations vs density}
\label{sec:SystemSizes}

\begin{figure*}[]
\includegraphics[width=0.75\textwidth]{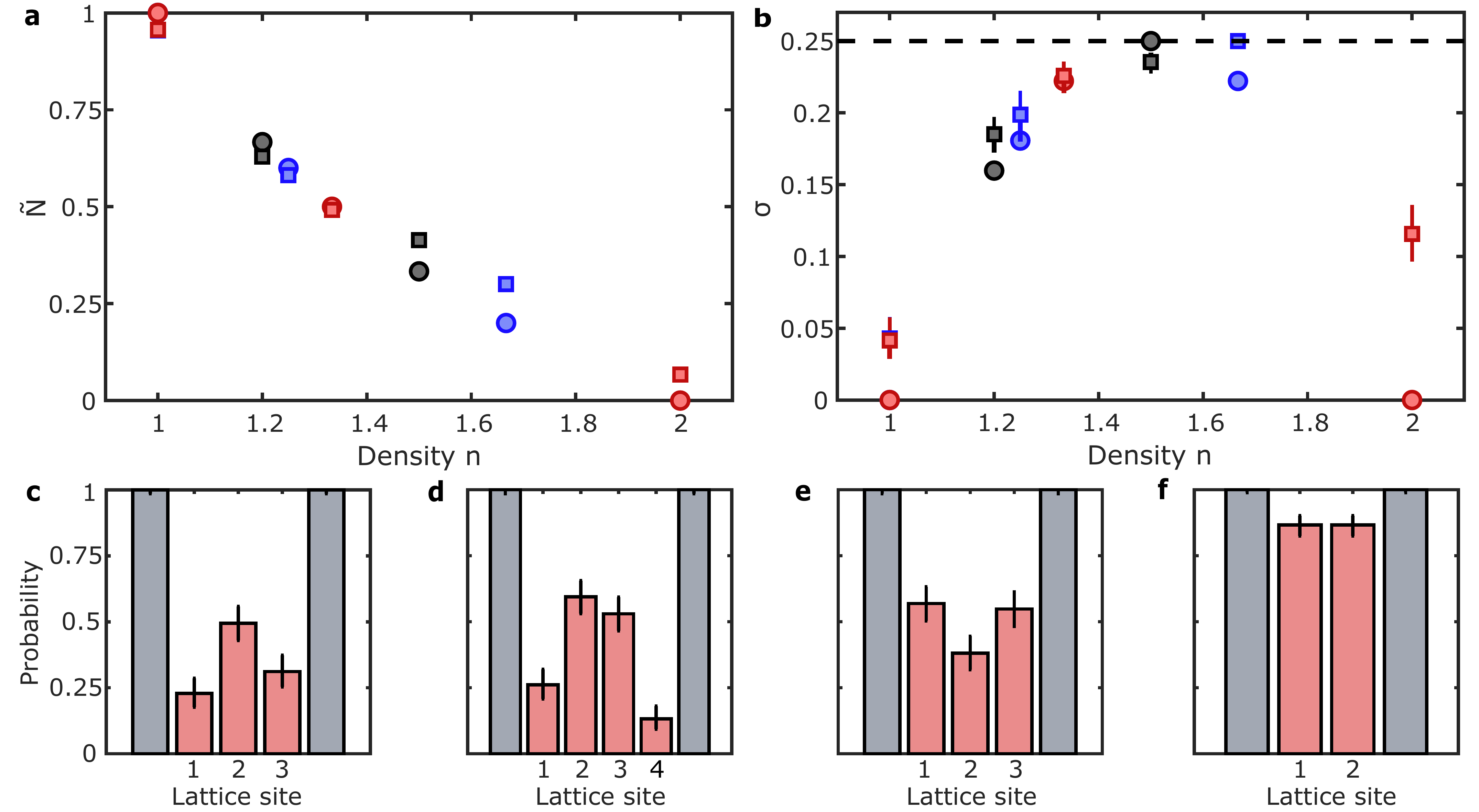}
\caption{\label{fig:DensityScan}\textbf{Atom number, variance, and site occupation  of commensurate and incommensurate systems. a}, Observed atom number normalised by the initial number of atoms, $\tilde{N}$, vs density, $n$, for systems of 6 atoms prepared on 6 sites dynamically compressed to 5 and 4 sites (black), 5 atoms on 5 sites compressed to 4 and 3 sites (blue), and 4 atoms on 4 sites compressed to 3 and 2 sites (red).  Experimental values are shown as squares, theoretical ones as circles. The statistical errors of the experimental values are smaller than the size of the datapoints. \textbf {b}, Atom number variance, using the same densities as in \textbf{a}. \textbf{c-f}, Site-resolved probability to detect an empty lattice site with increasing density, for 4 atoms on 3 sites, 6 atoms on 4 sites, 5 atoms on 3 sites and 4 atoms on 2 sites, respectively. Error bars in \textbf{b-f} are the 95\% Clopper-Pearson confidence intervals.}
\end{figure*}

The specific number of atoms and sites available in an incommensurate system can lead to non-trivial ground-state occupations that depend on the system size. We have so far considered 5 atoms on 4 sites, and we now compare different incommensurate states with 4, 5 and 6 particles to a Mott insulator with unit or double site occupancy in the strongly interacting regime. For each 1D system, we evaluate as a function of the average particle density, $n$, the detected atom number normalised by the number of sites before dynamic compression, $\tilde{N}$, (Fig.~\ref{fig:DensityScan}a). We also calculate the variance, $\sigma$, from the mean atom number parity (Methods) (Fig.~\ref{fig:DensityScan}b).
As expected, we observe Mott-insulating states with $n=1$ (4 atoms on 4 sites), where $\tilde{N} \approx 1$ and $n=2$ (4 atoms on 2 sites), where $\tilde{N} \approx 0$, as doubly occupied sites are detected as empty sites due to light-assisted collisions. The observed atom number decreases with increasing density, in agreement with our numerical calculations for the ground states at $T=0$ (Methods) indicating adiabatic state preparation. The variance, $\sigma$,  which is a measure for the compressibility for short-range density fluctuations \cite{Rigol2009}, is lowest at integer densities in the Mott-insulating state. It attains its maximum value of $\sigma=0.25$  for non-integer densities  \cite{Sherson2010}, again in agreement with the numerical calculations at zero temperature (Fig.~\ref{fig:DensityScan}b).

\begin{figure}[h!]
\centering
\includegraphics[width=\linewidth]{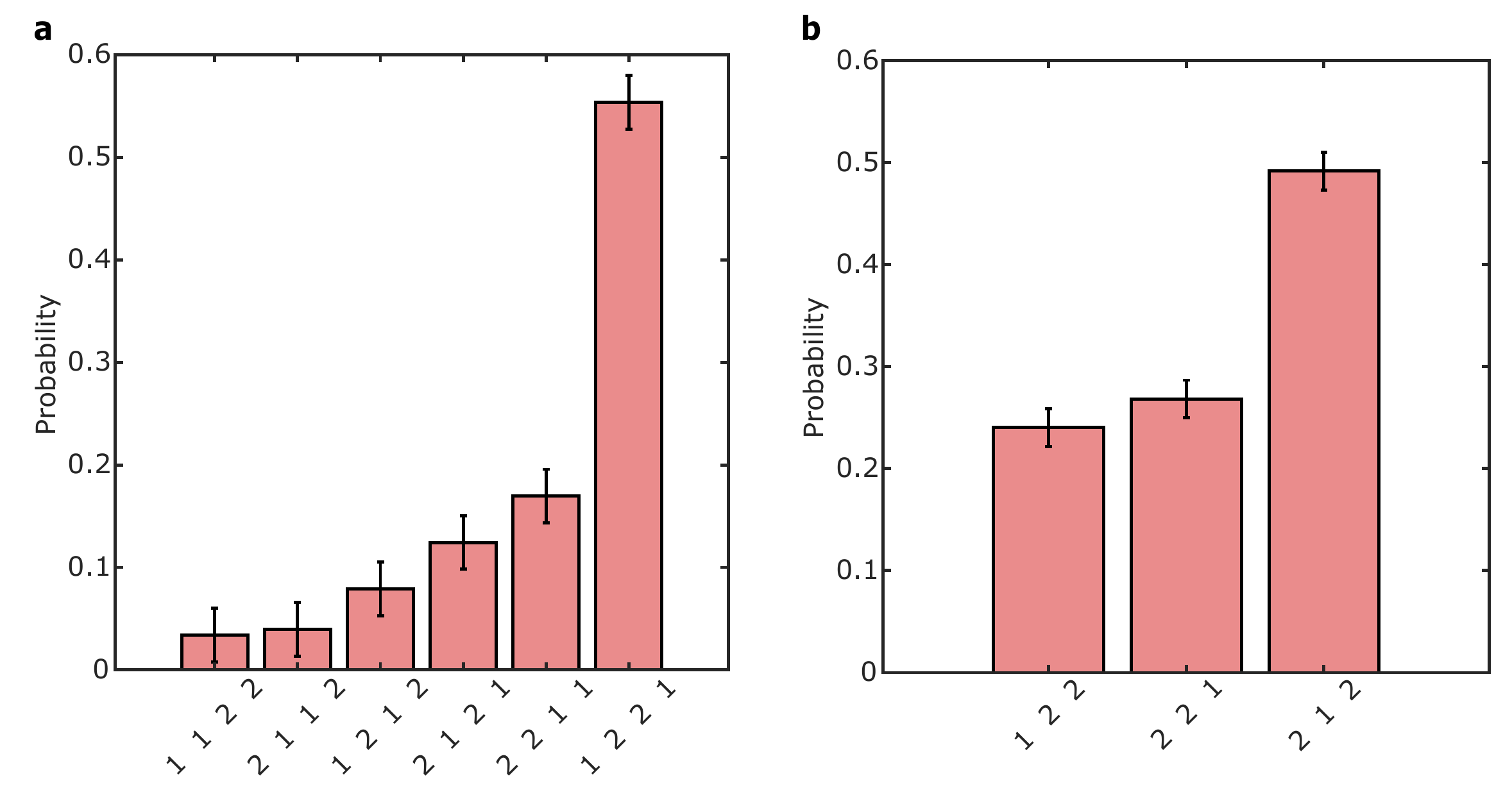}
\caption{\label{fig:stateselection}\textbf{State detection probabilities  for incommensurate systems with two additional particles.} \textbf{a}, Probability to detect the different number states for a system  with 6 atoms on 4 sites.  \textbf{b}, same for the system  with 5 atoms on 3 sites. }
\end{figure}

In the case of two additional atoms on a localised background, the symmetry of the system plays an important role. For 5 atoms on 3 sites ($n=5/3$),  it is energetically unfavourable for both additional atoms to be on the same lattice site. We observed that doubly occupied sites have a higher probability to be found on the outer sites compared to the central site (Fig.~\ref{fig:DensityScan}e). The state $\ket{2,1,2}$ is favourable (Supplementary Information, Fig. \,\ref{fig:stateselection}b), as it couples to both $\ket{1,2,2}$ and $\ket{2,2,1}$, reducing the kinetic energy of the ground state,
which is analytically given by $\frac{1}{\sqrt{2}}\ket{2,1,2} + \frac{1}{2}\ket{1,2,2} + \frac{1}{2}\ket{2,2,1}$ in the limit of $U/J\rightarrow\infty$ (Supplementary Information).
This state is robust against the presence of a weak harmonic confining potential and the small offsets from the adjacent  potential walls (Supplementary Information). This is in contrast to the state with 6 atoms on 4 sites ($n=6/4$), for which we observe the additional atoms mostly on the inner two sites (Fig.~\ref{fig:DensityScan}d).
Specifically, out of the systems post-selected to have two empty sites, we observe the empty sites (corresponding to sites with two atoms) next to each other in 76(7)\% of the cases. In 55(7)\% of the cases the two empty sites are in the centre, corresponding to the observation of state  $\ket{1,2,2,1}$ (Supplementary Information, Fig.\,\ref{fig:stateselection}a). Unlike the $n=5/3$ system, the $n=6/4$ system is very sensitive to additional potential offsets, such that  the inclusion of the harmonic confinement and wall potentials leads to a favoured occupation of the central sites, while in a perfect box potential the predicted density profile is flat. All our observations are well explained by the single-band Bose-Hubbard model, while being consistent with previous numerical calculations  beyond the single-band model \cite{Brouzos2010}. The inclusion of higher bands was shown to result in repulsion effects and fragmentation of the  on-site density. While such effects will be present here, their observation would require the ability to probe  the atomic wave function with sub-lattice-site resolution.

\subsection*{Particle mobility in a bias potential}

\begin{figure*}[]
\includegraphics[width= 0.80 \textwidth]{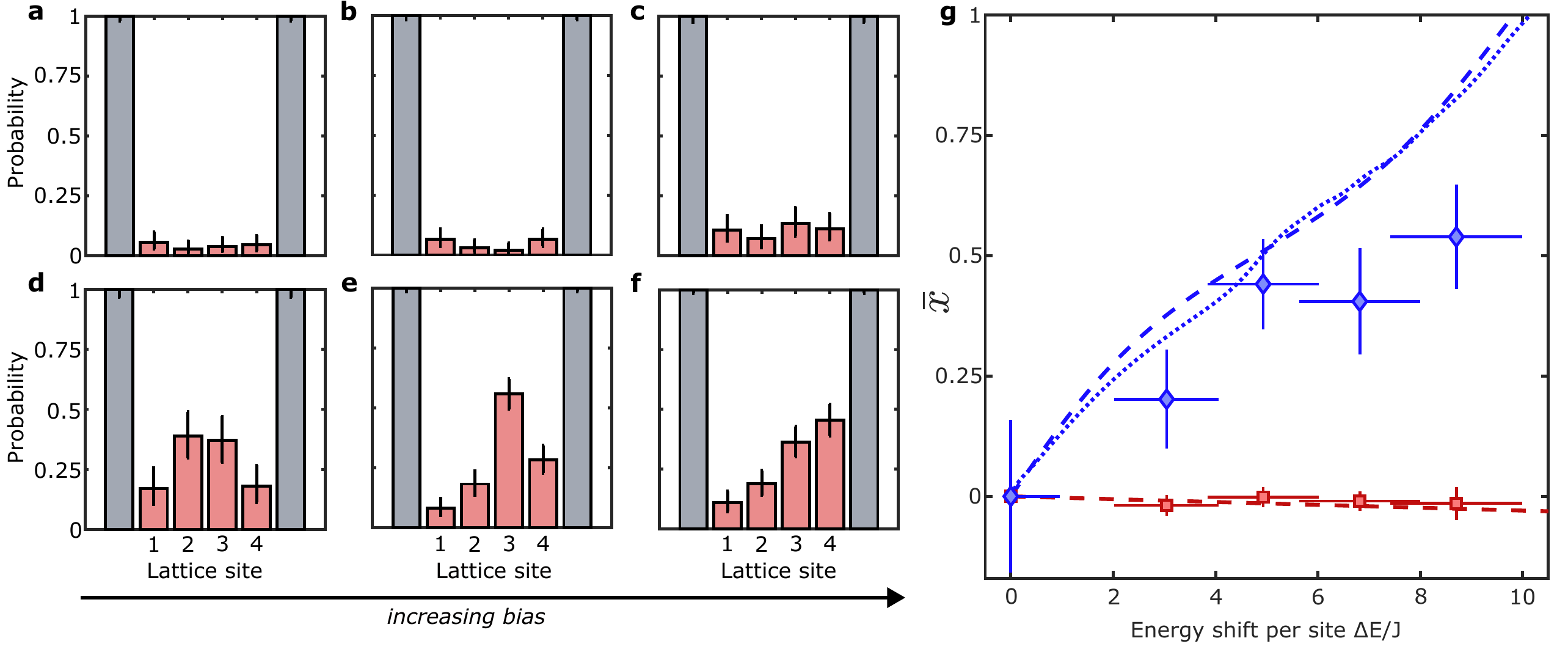}
\caption{\label{fig:Gradient}\textbf{Mott insulator and doped insulator in a bias potential. a-c,} Probability distributions of detecting an empty site in a commensurate system (four atoms on four sites) for maximum energy shifts per lattice site
$\Delta E/J=0(1), 5(1), 9(1)$ resulting from the bias potential. \textbf{d-f,} same distributions of an incommensurate system for the same potential (five atoms on four sites), where the empty site can result from the detection of a doublon. \textbf{g}, Centre-of-mass shift, $\bar{x}$, measured in lattice sites, relative to the original distribution as a function the energy shift $\Delta E$,  for a commensurate (red squares) and incommensurate (blue diamonds) system, together with the corresponding numerical simulations of the ground state (dashed lines) and the ensemble average  (dotted lines). Errors of  $\Delta E/J$ and $\bar x$ are calculated via  error propagation from the gradient calibration and from the  counting statistics, respectively. Each histogram  is obtained by averaging over $100-300$ 1D  systems.}
\end{figure*}

It is expected that doped insulators will behave differently from undoped ones when subject to external probes, e.g., when measuring their compressibility and particle mobility. To show this we investigate how the commensurate and incommensurate systems change when subject to a gradient potential of the form $\hat{H}_{g}= \Delta E\,\sum_{i=1}^N i\,\hat{n}_i$, where $\Delta E$ denotes the energy shift per lattice site and $\hat{n}_i$ the local number operator. We use a deep lattice, $V_x=16(1)\,E_{\rm r}$, $J/h = \unit[9(1)]{Hz}$, and a  magnetic bias field that is slowly increased within 500\,ms to maintain adiabaticity. When the bias field is applied to a commensurate system of four atoms on four sites, the  distribution of the the empty sites remains almost unchanged  (Fig.\,\ref{fig:Gradient}a - c) \cite{Gemelke2009}. In the incommensurate system of five atoms on four lattice sites,  the probability of finding the additional atom (detected as an empty site), is skewed in the direction of the force produced by the gradient (Fig.~\ref{fig:Gradient}d - f), showing that the doped insulator has a different and nonzero compressibility compared to the undoped state.  We quantify this  effect by computing the centre of mass, $\bar{x}$, of the histograms in Figs.\,\ref{fig:Gradient}a to \ref{fig:Gradient}f as a function of $\Delta E$. While  there is no change of the centre of mass for the commensurate system (Fig.~\ref{fig:Gradient}g, red squares),  for the incommensurate system (Fig.~\ref{fig:Gradient}g, blue diamonds),  $\bar{x}$ increases with $\Delta E$, showing that the incommensurate (doped) system is compressible  \cite{Ho2010,Busley2022}. The centre-of-mass shift  is sensitive to the specific shape of the potential barriers and the harmonic confinement (Supplementary Information), which is accounted for by the numerical simulation of the system dynamics.

%
%
\subsection*{\label{sec:level1}Discussion}
We have studied the effects of commensurability in  one-dimensional bosonic quantum systems. Key to this is our ability to produce engineered dynamical light potentials at the scale of single lattice sites. Starting from a  commensurate filling with a known atom number between static potential barriers, we  moved the  barriers  to change  the number of available lattice sites, producing incommensurate systems. To characterise  our degree of control of the state preparation,  we characterised these systems by measuring  the occurrence of holes and doublons from strong to weak interactions. For incommensurate systems, featuring  delocalised atoms on a localised background, we observed non-trivial site  occupation probabilities, in agreement with our numerical calculations.
Studying the spatial distribution of our systems in a potential gradient, we observed particle mobility and compressibility of  the incommensurate systems, while the commensurate ones remain in an incompressible  Mott-insulating state. To enable the study of larger systems, a weaker external confinement would be required, which can be achieved by programming a deconfining potential using the DMD. This would also allow us to study systems doped with  holes instead of particles.
Introducing disorder to incommensurate systems leads to a way to further explore the transitions between superfluid and Bose glass for few-boson systems and the effect on compressibility \cite{Fisher1989,Damski2003,Cai2010}. For ladder systems, control over the atom number can lead to further interesting effects including the realisation of a `rung Mott insulator',  predicted in a two-leg ladder with half filling \cite{Carrasquilla2011, Crepin2011}.  Our methods for generating dynamically controlled potentials can also be used to adiabatically prepare low entropy states with controlled incommensurability or doping in both bosonic and fermionic systems.

\section*{Methods}
\label{sec:Methods}
\subsection*{Experimental procedure}
\label{sec:Methods-Exp}

The experiment starts with a cloud of  $2\times 10^9$  $^{87}$Rb atoms in a magneto-optical trap (MOT) loaded from a  2D$^+$ MOT. The cloud is compressed by increasing the MOT's magnetic quadrupole field and cooled to $\unit[2]{{\mu}K}$ using Raman grey molasses cooling on the D2 line \cite{greymolasses}. We load about $1\times 10^8$ atoms into a crossed optical dipole trap (CODT), formed by two laser beams at 1070\,nm wavelength and  200\,W power, intersecting at a 17(1)$^\circ$ angle. The waist of the two laser beams are $425(5){\mu}$m.
After this, the atoms are loaded into the focus of a dimple trap with a waist of $47(5){\mu}$m. The beam is moved using a translation stage, transporting the atoms into the `science' chamber equipped with a high-resolution microscope objective  ($\mbox{NA}=0.69$). About $3\times 10^6$ atoms are transferred into another CODT that is formed by the optical lattice beams with the retro-reflected beams blocked,  before they  are cooled further using evaporative cooling and loaded into the vertical optical lattice. We use a position-dependent microwave transfer in a magnetic field gradient to create a  two-dimensional cloud of thermal atoms in a single anti-node of the vertical optical lattice \cite{Sherson2010}.

Then, a  dimple trap at 850\,nm wavelength is shone in through the microscope objective, while keeping the vertical lattice  depth at $V_z=20(1) E_r$. We use a magnetic quadrupole field to tilt the trap, such that we evaporatively cool the atoms to create a Bose-Einstein condensate. We now shine in the blue-detuned light shaped by the  DMD  to create the  repulsive potential barriers, before  the  two  horizontal optical lattice beams are turned on.  We detect the individual atoms via fluorescence imaging using the microscope objective, similar to previous works \cite{Sherson2010}.
To freeze the atom distribution prior to detection, the optical lattice depth is first increased  to $50(2) E_r$ in  $\unit[500]{\mu s}$ in all three axes, and then to $3000E_r$ in $\unit[2]{ms}$. The Lucy-Richardson algorithm is used to deconvolve the fluorescence images and reconstruct  the lattice occupation with high fidelity \cite{larooij2022}.  We estimate the overall detection fidelity to be $\approx 99\%$, limited by atom hopping and losses during imaging, and by  the image reconstruction fidelity.

\subsection*{Programmable dynamic light potentials}
\label{sec:Methods-DMD}
We use a DMD (ViALUX V-9001) to create the repulsive potential barriers using  blue-detuned light at \unit[666]{nm} wavelength from an amplified diode laser. The DMD image is projected onto the atoms by the high-resolution objective such that we can use   $18\times 18$ DMD pixels per lattice site to control our custom potentials.  A dedicated software allows us to specify initial and final positions of the potential barriers  in the reference frame of the optical lattice. We can also set the number of different patterns (frames) to be displayed on the DMD. An initial frame is displayed  during the system preparation, and after a trigger pulse, a sequence of frames  are displayed moving the barrier to reduce the system size by a discrete number of lattice sites. We program the DMD in `uninterrupted' mode to suppress the dark time between successive frames. The transition time  between frames is  $8\,\mu$s during which the mirrors are released and the next configuration of mirrors is switched on.

\begin{figure}[!t]
\includegraphics[width=0.8\columnwidth]{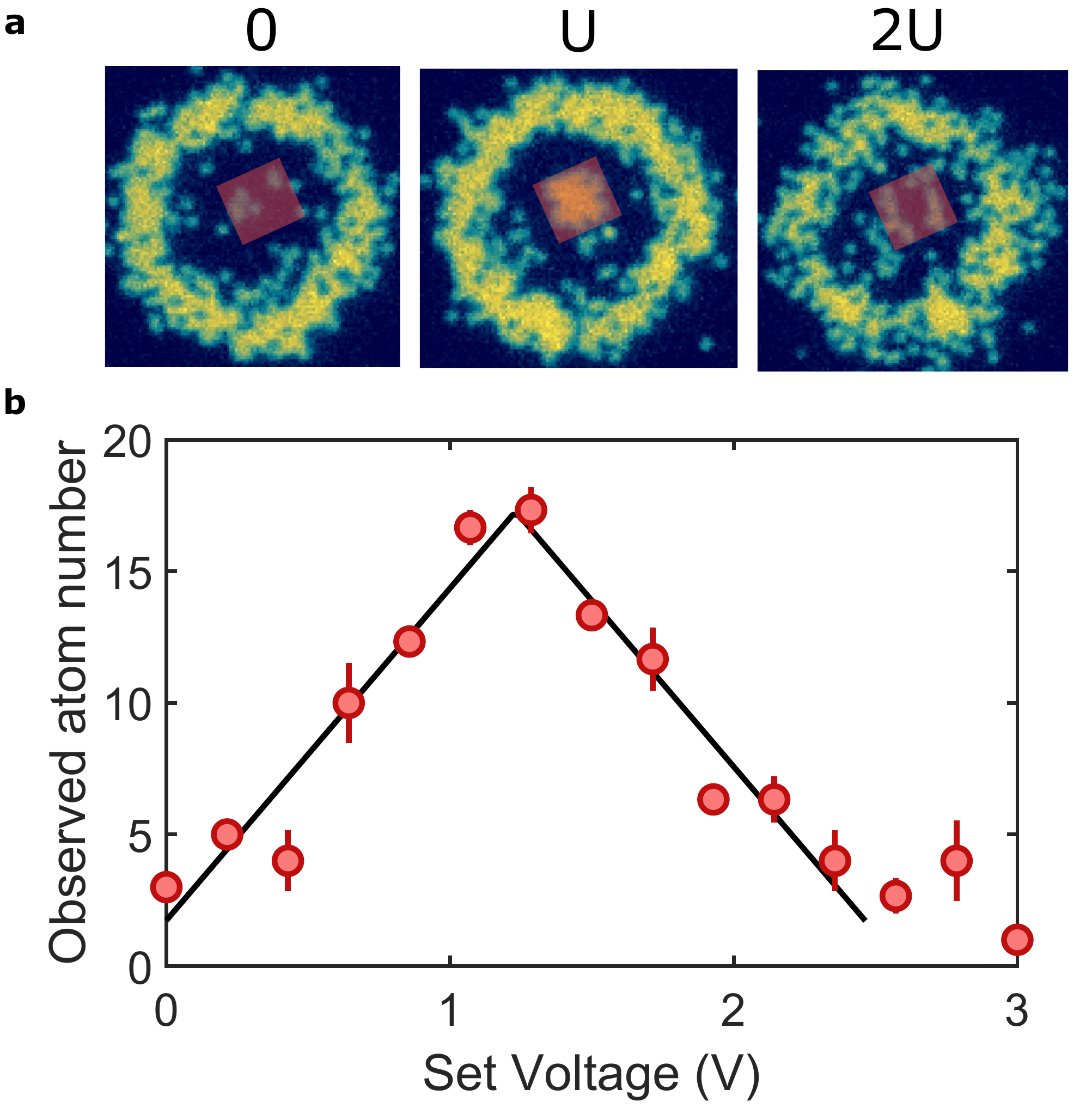}
\centering
\caption{\label{fig:PowerCalibration}\textbf{Calibration of the repulsive barriers. a}, Fluorescence image of atoms in a Mott insulating state with an inner $n=2$ shell (visible as mostly empty sites), illuminated with a square repulsive potential in a  region of $5 \times 4$ sites, indicated  by the red square. For the  three images, three different intensities of the 666\,nm light were used, corresponding to potential heights of 0, $U$, and $2U$. The potential depths of the optical lattices were  $V_x=V_y= 20(1)\,E_{r}$ and $V_z =35(5)\,E_{r}$, such that $U/h=\unit[940(100)]{Hz}$. \textbf{b},  Observed atom number within the region highlighted by the red square in \textbf{a}, as function of the set voltage of the  laser intensity regulation. Each data point is obtained by averaging the counted atom number in three images, the error bars are standard error. The peak of the graph represents a potential height $U$ at set voltage of 1.23(8)V.
}
\end{figure}

We observe a  phase drift of the optical lattices \cite{Weitenberg2011}, from one realisation of the experiment with respect to the next one, resulting in position shift of approximately $0.05\,a_{\rm l}$ on average, where $a_{\rm l} = \lambda/2$ is the lattice spacing. This position shift is  measured by fitting the position of single atoms in the fluorescence image. We use this information to shift the position of the DMD pattern for the next measurement to follow the phase drift. In this way the repulsive potential stays in the same place with respect to the optical lattice, with an estimated deviation of $<0.05$ times the lattice spacing.

To calibrate the intensity of the 666\,nm light, a repulsive potential is projected onto the centre $5\times 4$ sites of an $n=2$ shell in a Mott insulating state (Fig.~\ref{fig:PowerCalibration}a). The number of atoms observed in the region as a function of the set voltage for the laser intensity regulation shows a peak (Fig.~\ref{fig:PowerCalibration}b) when  the light shift caused by the repulsive light potential is equal to $U$. In this case, all sites covered by the repulsive potential have  single occupancy and form an $n=1$ Mott insulating shell \cite{Hirthe2023}. When increasing the light intensity of the repulsive potential further such that the light shift equals  $2U$, we eventually see no atoms in the region once again.

We verified that compressing the quantum gas in the superfluid regime using the dynamic DMD potential does not result in significant atom loss when using a `frame duration' of 10~ms. To quantify the atom losses, we have reversed the position of the repulsive potential barriers to its  original position over the same timescale as for the compression, then transferred the system back into the Mott-insulating state. We found that with static barriers we detected on average 4.81(5) atoms on five sites, while when moving and reversing the potential barriers we detected on average 4.64(7) atoms.

\subsection*{Numerical simulation of system dynamics}
For strongly interacting systems towards the Tonks-Girardeau limit with filling above unity, it is known that higher bands need to be accounted for as the on-site repulsion results in a fragmentation of the density of the particles in the ground state within individual sites \cite{Brouzos2010}. However, for the case considered here, the quantum gas microscope can resolve between single sites and not for the particle density within a site. Therefore, we consider the single-band (lowest energy) Bose-Hubbard model to model the atoms in the one-dimensional optical lattices. The Hamiltonian is \cite{Fisher1989,Jaksch1998}
\begin{equation}
    \hat{H}=-J\sum_{i=1}^{N}\left(\hat{a}_i^\dagger\hat{a}_{i+1}+h.c.\right)+\frac{U}{2}\sum_{i=1}^{N}\hat{n}_i(\hat{n}_i-1)+\sum_{i=1}^{N} \epsilon_i \hat{n}_i,\label{eq:BHM}
\end{equation}
where $N$ is the number of sites, $\hat{a}_i^\dagger$ and $\hat{a}_i$ are  the bosonic creation and annihilation operators respectively, $\hat{n}_i=\hat{a}_i^\dagger\hat{a}_i$ is the number operator, $J$ is the tunneling strength between neighbouring lattice sites, $U$ is the on-site interaction energy, and $\epsilon_i$ is the local energy shift due to an external potential. We calculate $J$ and $U$ from the Wannier functions, while $\epsilon_i$ accounts for the weak harmonic confinement and the impact of the wall potential, taking into account the calibrated height of the potential barriers and the point-spread function of the microscope.

We emulate the dynamics of the experiment numerically, accounting for the full Hilbert space of the finite lattice with fixed particle number in all cases. The state is initialised in the Mott insulator of the lattice with commensurate filling  and we numerically implement the same protocol for higher filling factors. For the ramps between deep and shallow lattice potentials, we evolve the system through the implementation of the unitary evolution operator for discrete time steps, with a maximum error for individual steps of $10^{-6}$.
We simulate the discrete steps of the potentials caused by the discrete frames of the DMD pattern when moving the barriers. To account for an initial thermal distribution of the state, we evolve each initial eigenstate individually and calculate  the final non-zero temperature state as the sum of the evolved states weighted by the Boltzmann distribution. Overall, we find good agreement between the zero temperature numerical results and the experiment.
The imaging quench to a deep lattice was simulated across a range of starting $J/U$ values for both commensurate and incommensurate densities to ensure that it does not introduce non-adiabatic effects and results in a frozen density profile.
As the same imaging quench is  used in each of the experimental realisations, we exclude it from the simulation.

There is a small probability that the initial one-dimensional system of the experiment has one more or one less atom than the number of lattice sites.  We have simulated the impact of this non-perfect state preparation to confirm that the observation of additional empty sites is not due to excitations from non-adiabatic effects, as these would not be captured by our numerical protocol with fixed atom number. We  emulate the imperfect preparation by simulating the case of one additional atom and one less atom. We then combine the results with those assuming a perfect initial state preparation, to mimic what we observe experimentally using the post-selection process.

\subsection*{Definition and calculation of observables}
The numerical simulations compute the full wave function, and for comparison with the experimental results we calculate the local parity operator, $\hat s_i$, of the $i$th site,
\begin{equation}\label{eq:localParity}
\hat s_i = \frac{1}{2}\left[ \left(-1\right)^{\hat{n}_i-1} + 1\right],
\end{equation}
with the local number operator $ \hat n_i$. We use the lattice occupation to measure the parity of the atom number a single lattice site, $s_i=\langle \hat s_i \rangle$, and the mean atom number parity on $M$ lattice sites, $\bar{n}= \sum_{i=1}^{M} s_i/M$, where $M$ is given by the size of the 1D systems multiplied by the number of realisations. From this, we calculate the variance of the atom number $\sigma = \bar n (1-\bar n)$, shown in Fig.\,\ref{fig:DensityScan}b.

For the datasets in Fig.~\ref{fig:Gradient}g, we compute the centre of mass  in the case of the incommensurate system (5 atoms on 4 sites) using $\bar{x}=\sum_{i=1}^{4}{i\, w_i}/\sum_{i=1}^{4}{w_i}$, where $w_i$ is the probability to find an empty site (i.e., the extra atom on a doubly occupied site) on lattice site $i$.  To match this for the commensurate system, we use $\bar{w}_i=1-w_i$ instead of $w_i$, such that for the histograms of both commensurate and incommensurate systems the centre of mass shift of the atoms is calculated.

\smallskip

\subsection*{Calibration of bias potential}
\begin{figure}[!h]
\includegraphics[width=0.8\linewidth]{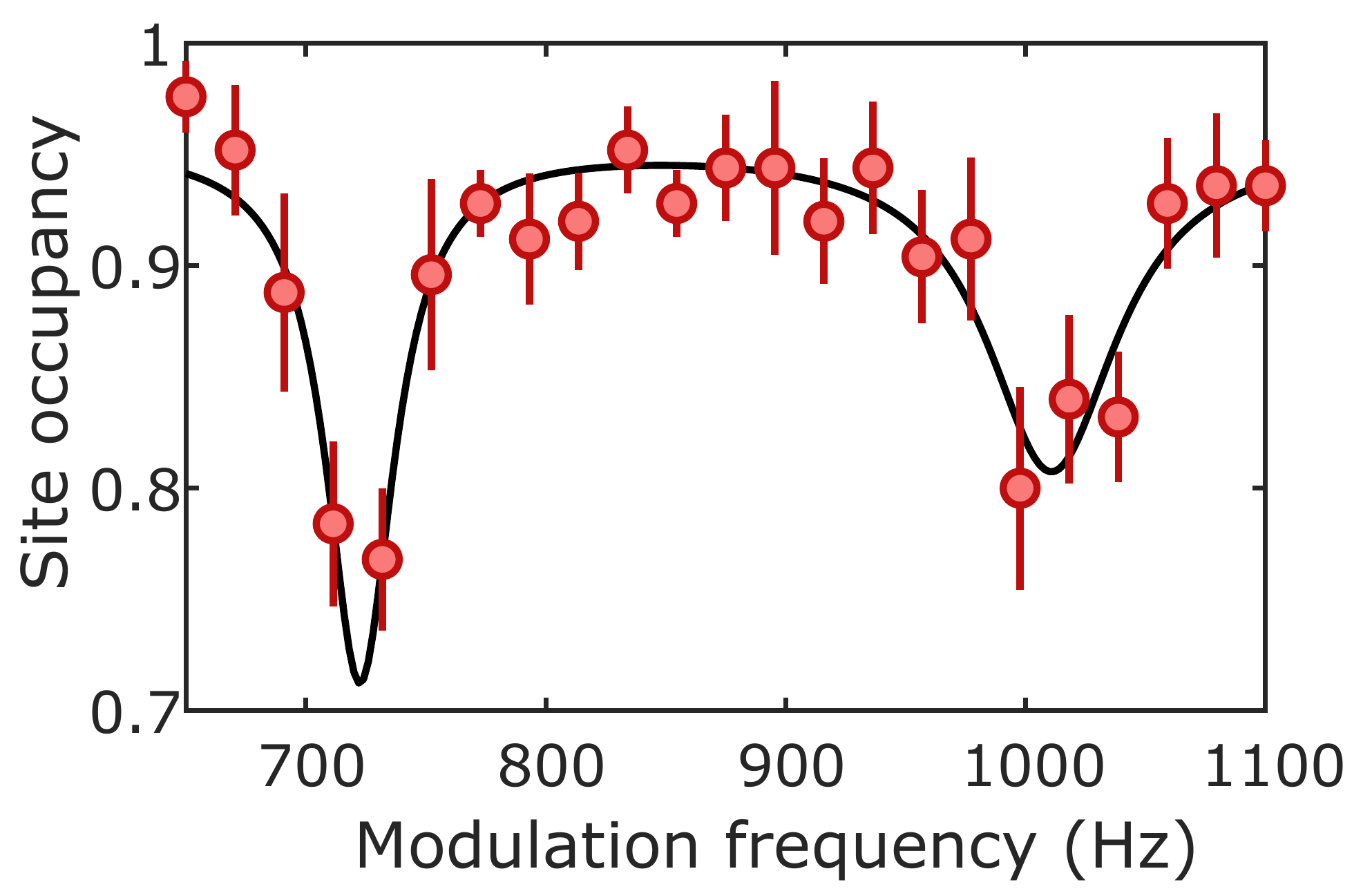}
\centering
\caption{\label{fig:gradientcalibration}\textbf{Calibration of the magnetic gradient field}. Observed  occupancy within the centre $5\times 5$ sites of a $n=1$ Mott Insulator at $U/h = \unit[830(100)]{Hz}$ versus the modulation frequency. The line is a fit with a double Lorentzian to obtain the peak positions. Each data point is obtained by averaging over the atom number counted of  five images. The error bars are the standard error.}
\end{figure}

To calibrate the magnetic field used for the bias potential, we follow a method used in previous studies \cite{Simon2011,Ma2011}. Starting with an $n=1$ Mott Insulator, we modulate the power of the $x$-lattice beams by 20\,\% for \unit[30]{ms}. When the frequency of the modulation matches the interaction energy, $U$, the atoms tunnel to already occupied sites which leads to an increase in holes and doubly-occupied sites. When a magnetic field gradient is applied, causing an energy offset per site of $\Delta E$, the atoms tunnel when the frequency of the modulation matches $U \pm \Delta E$. The number of atoms counted in a central region as a function of $\Delta E$ shows two inverted peaks (Fig.~\ref{fig:gradientcalibration}), the position which we identify by fitting  a double Lorentzian, yielding $U \pm \Delta E$. We repeated this  measurement for different field gradients, and fit a linear regression model from which we obtain the error bars for $\Delta E/J$ shown in Fig.~\ref{fig:Gradient}g.

\section*{Acknowledgements}
We acknowledge support by the Engineering and Physical Sciences Research Council (EPSRC) through the Programme Grant DesOEQ [grant number EP/P009565/1], the Quantum Technology Hub in Quantum Computing and Simulation [EP/T001062/1], the New Investigator Grant [EP/T027789/1], and the Doctoral Training Partnership grants for CP [EP/T517938/1] and LK
[EP/W524670/1]. We thank Ilian Despard, Andrés Ulibarrena, Harikesh Ranganath and Maximilan Ammenwerth for their work during  the early stages of the experimental setup.

\bibliography{ComIncomPaperRefs}

\begin{thebibliography}{50}%
\makeatletter
\providecommand \@ifxundefined [1]{%
 \@ifx{#1\undefined}
}%
\providecommand \@ifnum [1]{%
 \ifnum #1\expandafter \@firstoftwo
 \else \expandafter \@secondoftwo
 \fi
}%
\providecommand \@ifx [1]{%
 \ifx #1\expandafter \@firstoftwo
 \else \expandafter \@secondoftwo
 \fi
}%
\providecommand \natexlab [1]{#1}%
\providecommand \enquote  [1]{``#1''}%
\providecommand \bibnamefont  [1]{#1}%
\providecommand \bibfnamefont [1]{#1}%
\providecommand \citenamefont [1]{#1}%
\providecommand \href@noop [0]{\@secondoftwo}%
\providecommand \href [0]{\begingroup \@sanitize@url \@href}%
\providecommand \@href[1]{\@@startlink{#1}\@@href}%
\providecommand \@@href[1]{\endgroup#1\@@endlink}%
\providecommand \@sanitize@url [0]{\catcode `\\12\catcode `\$12\catcode
  `\&12\catcode `\#12\catcode `\^12\catcode `\_12\catcode `\%12\relax}%
\providecommand \@@startlink[1]{}%
\providecommand \@@endlink[0]{}%
\providecommand \url  [0]{\begingroup\@sanitize@url \@url }%
\providecommand \@url [1]{\endgroup\@href {#1}{\urlprefix }}%
\providecommand \urlprefix  [0]{URL }%
\providecommand \Eprint [0]{\href }%
\providecommand \doibase [0]{https://doi.org/}%
\providecommand \selectlanguage [0]{\@gobble}%
\providecommand \bibinfo  [0]{\@secondoftwo}%
\providecommand \bibfield  [0]{\@secondoftwo}%
\providecommand \translation [1]{[#1]}%
\providecommand \BibitemOpen [0]{}%
\providecommand \bibitemStop [0]{}%
\providecommand \bibitemNoStop [0]{.\EOS\space}%
\providecommand \EOS [0]{\spacefactor3000\relax}%
\providecommand \BibitemShut  [1]{\csname bibitem#1\endcsname}%
\let\auto@bib@innerbib\@empty
\bibitem [{\citenamefont {Bakr}\ \emph {et~al.}(2010)\citenamefont {Bakr},
  \citenamefont {Peng}, \citenamefont {Tai}, \citenamefont {Ma}, \citenamefont
  {Simon}, \citenamefont {Gillen}, \citenamefont {F{\"{o}}lling}, \citenamefont
  {Pollet},\ and\ \citenamefont {Greiner}}]{Bakr2010}%
  \BibitemOpen
  \bibfield  {author} {\bibinfo {author} {\bibfnamefont {W.~S.}\ \bibnamefont
  {Bakr}}, \bibinfo {author} {\bibfnamefont {A.}~\bibnamefont {Peng}}, \bibinfo
  {author} {\bibfnamefont {M.~E.}\ \bibnamefont {Tai}}, \bibinfo {author}
  {\bibfnamefont {R.}~\bibnamefont {Ma}}, \bibinfo {author} {\bibfnamefont
  {J.}~\bibnamefont {Simon}}, \bibinfo {author} {\bibfnamefont {J.~I.}\
  \bibnamefont {Gillen}}, \bibinfo {author} {\bibfnamefont {S.}~\bibnamefont
  {F{\"{o}}lling}}, \bibinfo {author} {\bibfnamefont {L.}~\bibnamefont
  {Pollet}},\ and\ \bibinfo {author} {\bibfnamefont {M.}~\bibnamefont
  {Greiner}},\ }\bibfield  {title} {\bibinfo {title} {{Probing the
  superfluid-to-Mott insulator transition at the single-atom level}},\ }\href
  {https://doi.org/10.1126/science.1192368} {\bibfield  {journal} {\bibinfo
  {journal} {Science}\ }\textbf {\bibinfo {volume} {329}},\ \bibinfo {pages}
  {547} (\bibinfo {year} {2010})}\BibitemShut {NoStop}%
\bibitem [{\citenamefont {Sherson}\ \emph {et~al.}(2010)\citenamefont
  {Sherson}, \citenamefont {Weitenberg}, \citenamefont {Endres}, \citenamefont
  {Cheneau}, \citenamefont {Bloch},\ and\ \citenamefont {Kuhr}}]{Sherson2010}%
  \BibitemOpen
  \bibfield  {author} {\bibinfo {author} {\bibfnamefont {J.~F.}\ \bibnamefont
  {Sherson}}, \bibinfo {author} {\bibfnamefont {C.}~\bibnamefont {Weitenberg}},
  \bibinfo {author} {\bibfnamefont {M.}~\bibnamefont {Endres}}, \bibinfo
  {author} {\bibfnamefont {M.}~\bibnamefont {Cheneau}}, \bibinfo {author}
  {\bibfnamefont {I.}~\bibnamefont {Bloch}},\ and\ \bibinfo {author}
  {\bibfnamefont {S.}~\bibnamefont {Kuhr}},\ }\bibfield  {title} {\bibinfo
  {title} {{Single-atom-resolved fluorescence imaging of an atomic Mott
  insulator}},\ }\href {https://doi.org/10.1038/nature09378} {\bibfield
  {journal} {\bibinfo  {journal} {Nature}\ }\textbf {\bibinfo {volume} {467}},\
  \bibinfo {pages} {68} (\bibinfo {year} {2010})}\BibitemShut {NoStop}%
\bibitem [{\citenamefont {Gross}\ and\ \citenamefont {Bakr}(2021)}]{Gross2021}%
  \BibitemOpen
  \bibfield  {author} {\bibinfo {author} {\bibfnamefont {C.}~\bibnamefont
  {Gross}}\ and\ \bibinfo {author} {\bibfnamefont {W.~S.}\ \bibnamefont
  {Bakr}},\ }\bibfield  {title} {\bibinfo {title} {Quantum gas microscopy for
  single atom and spin detection},\ }\href
  {https://doi.org/10.1038/s41567-021-01370-5} {\bibfield  {journal} {\bibinfo
  {journal} {Nat. Phys.}\ }\textbf {\bibinfo {volume} {17}},\ \bibinfo {pages}
  {1316} (\bibinfo {year} {2021})}\BibitemShut {NoStop}%
\bibitem [{\citenamefont {Lewenstein}\ \emph {et~al.}(2007)\citenamefont
  {Lewenstein}, \citenamefont {Sanpera}, \citenamefont {Ahufinger},
  \citenamefont {Damski}, \citenamefont {Sen(De)},\ and\ \citenamefont
  {Sen}}]{Lewenstein2007}%
  \BibitemOpen
  \bibfield  {author} {\bibinfo {author} {\bibfnamefont {M.}~\bibnamefont
  {Lewenstein}}, \bibinfo {author} {\bibfnamefont {A.}~\bibnamefont {Sanpera}},
  \bibinfo {author} {\bibfnamefont {V.}~\bibnamefont {Ahufinger}}, \bibinfo
  {author} {\bibfnamefont {B.}~\bibnamefont {Damski}}, \bibinfo {author}
  {\bibfnamefont {A.}~\bibnamefont {Sen(De)}},\ and\ \bibinfo {author}
  {\bibfnamefont {U.}~\bibnamefont {Sen}},\ }\bibfield  {title} {\bibinfo
  {title} {Ultracold atomic gases in optical lattices: mimicking condensed
  matter physics and beyond},\ }\href
  {https://doi.org/10.1080/00018730701223200} {\bibfield  {journal} {\bibinfo
  {journal} {Advances in Physics}\ }\textbf {\bibinfo {volume} {56}},\ \bibinfo
  {pages} {243} (\bibinfo {year} {2007})}\BibitemShut {NoStop}%
\bibitem [{\citenamefont {Bloch}\ \emph {et~al.}(2008)\citenamefont {Bloch},
  \citenamefont {Dalibard},\ and\ \citenamefont {Zwerger}}]{Bloch2008}%
  \BibitemOpen
  \bibfield  {author} {\bibinfo {author} {\bibfnamefont {I.}~\bibnamefont
  {Bloch}}, \bibinfo {author} {\bibfnamefont {J.}~\bibnamefont {Dalibard}},\
  and\ \bibinfo {author} {\bibfnamefont {W.}~\bibnamefont {Zwerger}},\
  }\bibfield  {title} {\bibinfo {title} {Many-body physics with ultracold
  gases},\ }\href {https://doi.org/10.1103/RevModPhys.80.885} {\bibfield
  {journal} {\bibinfo  {journal} {Rev. Mod. Phys.}\ }\textbf {\bibinfo {volume}
  {80}},\ \bibinfo {pages} {885} (\bibinfo {year} {2008})}\BibitemShut
  {NoStop}%
\bibitem [{\citenamefont {Gross}\ and\ \citenamefont
  {Bloch}(2017)}]{Gross2017}%
  \BibitemOpen
  \bibfield  {author} {\bibinfo {author} {\bibfnamefont {C.}~\bibnamefont
  {Gross}}\ and\ \bibinfo {author} {\bibfnamefont {I.}~\bibnamefont {Bloch}},\
  }\bibfield  {title} {\bibinfo {title} {Quantum simulations with ultracold
  atoms in optical lattices},\ }\href {https://doi.org/10.1126/science.aal3837}
  {\bibfield  {journal} {\bibinfo  {journal} {Science}\ }\textbf {\bibinfo
  {volume} {357}},\ \bibinfo {pages} {995} (\bibinfo {year}
  {2017})}\BibitemShut {NoStop}%
\bibitem [{\citenamefont {Sch{\"{a}}fer}\ \emph {et~al.}(2020)\citenamefont
  {Sch{\"{a}}fer}, \citenamefont {Fukuhara}, \citenamefont {Sugawa},
  \citenamefont {Takasu},\ and\ \citenamefont {Takahashi}}]{Schafer2020}%
  \BibitemOpen
  \bibfield  {author} {\bibinfo {author} {\bibfnamefont {F.}~\bibnamefont
  {Sch{\"{a}}fer}}, \bibinfo {author} {\bibfnamefont {T.}~\bibnamefont
  {Fukuhara}}, \bibinfo {author} {\bibfnamefont {S.}~\bibnamefont {Sugawa}},
  \bibinfo {author} {\bibfnamefont {Y.}~\bibnamefont {Takasu}},\ and\ \bibinfo
  {author} {\bibfnamefont {Y.}~\bibnamefont {Takahashi}},\ }\bibfield  {title}
  {\bibinfo {title} {Tools for quantum simulation with ultracold atoms in
  optical lattices},\ }\href {https://doi.org/10.1038/s42254-020-0195-3}
  {\bibfield  {journal} {\bibinfo  {journal} {Nature Reviews Physics}\ }\textbf
  {\bibinfo {volume} {2}},\ \bibinfo {pages} {411} (\bibinfo {year}
  {2020})}\BibitemShut {NoStop}%
\bibitem [{\citenamefont {Kaufman}\ \emph {et~al.}(2016)\citenamefont
  {Kaufman}, \citenamefont {Tai}, \citenamefont {Lukin}, \citenamefont
  {Rispoli}, \citenamefont {Schittko}, \citenamefont {Preiss},\ and\
  \citenamefont {Greiner}}]{Kaufman2016}%
  \BibitemOpen
  \bibfield  {author} {\bibinfo {author} {\bibfnamefont {A.}~\bibnamefont
  {Kaufman}}, \bibinfo {author} {\bibfnamefont {E.~M.}\ \bibnamefont {Tai}},
  \bibinfo {author} {\bibfnamefont {A.}~\bibnamefont {Lukin}}, \bibinfo
  {author} {\bibfnamefont {M.}~\bibnamefont {Rispoli}}, \bibinfo {author}
  {\bibfnamefont {R.}~\bibnamefont {Schittko}}, \bibinfo {author}
  {\bibfnamefont {P.~M.}\ \bibnamefont {Preiss}},\ and\ \bibinfo {author}
  {\bibfnamefont {M.}~\bibnamefont {Greiner}},\ }\bibfield  {title} {\bibinfo
  {title} {{Quantum thermalization through entanglement in an isolated
  many-body system}},\ }\href {https://doi.org/10.1126/science.aaf6725}
  {\bibfield  {journal} {\bibinfo  {journal} {Science}\ }\textbf {\bibinfo
  {volume} {353}},\ \bibinfo {pages} {794} (\bibinfo {year}
  {2016})}\BibitemShut {NoStop}%
\bibitem [{\citenamefont {Lukin}\ \emph {et~al.}(2019)\citenamefont {Lukin},
  \citenamefont {Rispoli}, \citenamefont {Schittko}, \citenamefont {Tai},
  \citenamefont {Kaufman}, \citenamefont {Choi}, \citenamefont {Khemani},
  \citenamefont {L{\'e}onard},\ and\ \citenamefont {Greiner}}]{Lukin2019}%
  \BibitemOpen
  \bibfield  {author} {\bibinfo {author} {\bibfnamefont {A.}~\bibnamefont
  {Lukin}}, \bibinfo {author} {\bibfnamefont {M.}~\bibnamefont {Rispoli}},
  \bibinfo {author} {\bibfnamefont {R.}~\bibnamefont {Schittko}}, \bibinfo
  {author} {\bibfnamefont {E.~M.}\ \bibnamefont {Tai}}, \bibinfo {author}
  {\bibfnamefont {A.}~\bibnamefont {Kaufman}}, \bibinfo {author} {\bibfnamefont
  {S.}~\bibnamefont {Choi}}, \bibinfo {author} {\bibfnamefont {V.}~\bibnamefont
  {Khemani}}, \bibinfo {author} {\bibfnamefont {J.}~\bibnamefont
  {L{\'e}onard}},\ and\ \bibinfo {author} {\bibfnamefont {M.}~\bibnamefont
  {Greiner}},\ }\bibfield  {title} {\bibinfo {title} {{Probing entanglement in
  a many-body-localized system}},\ }\href
  {https://doi.org/10.1126/science.aau0818} {\bibfield  {journal} {\bibinfo
  {journal} {Science}\ }\textbf {\bibinfo {volume} {364}},\ \bibinfo {pages}
  {256} (\bibinfo {year} {2019})}\BibitemShut {NoStop}%
\bibitem [{\citenamefont {Mazurenko}\ \emph {et~al.}(2017)\citenamefont
  {Mazurenko}, \citenamefont {Chiu}, \citenamefont {Ji}, \citenamefont
  {Parsons}, \citenamefont {Kan{\'a}sz-Nagy}, \citenamefont {Schmidt},
  \citenamefont {Grusdt}, \citenamefont {Demler}, \citenamefont {Greif},\ and\
  \citenamefont {Greiner}}]{Mazurenko2017}%
  \BibitemOpen
  \bibfield  {author} {\bibinfo {author} {\bibfnamefont {A.}~\bibnamefont
  {Mazurenko}}, \bibinfo {author} {\bibfnamefont {C.~S.}\ \bibnamefont {Chiu}},
  \bibinfo {author} {\bibfnamefont {G.}~\bibnamefont {Ji}}, \bibinfo {author}
  {\bibfnamefont {M.~F.}\ \bibnamefont {Parsons}}, \bibinfo {author}
  {\bibfnamefont {M.}~\bibnamefont {Kan{\'a}sz-Nagy}}, \bibinfo {author}
  {\bibfnamefont {R.}~\bibnamefont {Schmidt}}, \bibinfo {author} {\bibfnamefont
  {F.}~\bibnamefont {Grusdt}}, \bibinfo {author} {\bibfnamefont
  {E.}~\bibnamefont {Demler}}, \bibinfo {author} {\bibfnamefont
  {D.}~\bibnamefont {Greif}},\ and\ \bibinfo {author} {\bibfnamefont
  {M.}~\bibnamefont {Greiner}},\ }\bibfield  {title} {\bibinfo {title} {A
  cold-atom fermi--hubbard antiferromagnet},\ }\href
  {https://doi.org/10.1038/nature22362} {\bibfield  {journal} {\bibinfo
  {journal} {Nature}\ }\textbf {\bibinfo {volume} {545}},\ \bibinfo {pages}
  {462} (\bibinfo {year} {2017})}\BibitemShut {NoStop}%
\bibitem [{\citenamefont {Sompet}\ \emph {et~al.}(2022)\citenamefont {Sompet},
  \citenamefont {Hirthe}, \citenamefont {Bourgund}, \citenamefont {Chalopin},
  \citenamefont {Bibo}, \citenamefont {Koepsell}, \citenamefont {Bojovi{\'c}},
  \citenamefont {Verresen}, \citenamefont {Pollmann}, \citenamefont {Salomon},
  \citenamefont {Gross}, \citenamefont {Hiler},\ and\ \citenamefont
  {Bloch}}]{Sompet2022}%
  \BibitemOpen
  \bibfield  {author} {\bibinfo {author} {\bibfnamefont {P.}~\bibnamefont
  {Sompet}}, \bibinfo {author} {\bibfnamefont {S.}~\bibnamefont {Hirthe}},
  \bibinfo {author} {\bibfnamefont {D.}~\bibnamefont {Bourgund}}, \bibinfo
  {author} {\bibfnamefont {T.}~\bibnamefont {Chalopin}}, \bibinfo {author}
  {\bibfnamefont {J.}~\bibnamefont {Bibo}}, \bibinfo {author} {\bibfnamefont
  {J.}~\bibnamefont {Koepsell}}, \bibinfo {author} {\bibfnamefont
  {P.}~\bibnamefont {Bojovi{\'c}}}, \bibinfo {author} {\bibfnamefont
  {R.}~\bibnamefont {Verresen}}, \bibinfo {author} {\bibfnamefont
  {F.}~\bibnamefont {Pollmann}}, \bibinfo {author} {\bibfnamefont
  {G.}~\bibnamefont {Salomon}}, \bibinfo {author} {\bibfnamefont
  {C.}~\bibnamefont {Gross}}, \bibinfo {author} {\bibfnamefont
  {T.}~\bibnamefont {Hiler}},\ and\ \bibinfo {author} {\bibfnamefont
  {I.}~\bibnamefont {Bloch}},\ }\bibfield  {title} {\bibinfo {title} {Realizing
  the symmetry-protected haldane phase in fermi--hubbard ladders},\ }\href
  {https://doi.org/10.1038/s41586-022-04688-z} {\bibfield  {journal} {\bibinfo
  {journal} {Nature}\ }\textbf {\bibinfo {volume} {606}},\ \bibinfo {pages}
  {484} (\bibinfo {year} {2022})}\BibitemShut {NoStop}%
\bibitem [{\citenamefont {Hirthe}\ \emph {et~al.}(2023)\citenamefont {Hirthe},
  \citenamefont {Chalopin}, \citenamefont {Bourgund}, \citenamefont
  {Bojovi\'c}, \citenamefont {Bohrdt}, \citenamefont {Demler}, \citenamefont
  {Grusdt}, \citenamefont {Bloch},\ and\ \citenamefont {Zeiher}}]{Hirthe2023}%
  \BibitemOpen
  \bibfield  {author} {\bibinfo {author} {\bibfnamefont {S.}~\bibnamefont
  {Hirthe}}, \bibinfo {author} {\bibfnamefont {T.}~\bibnamefont {Chalopin}},
  \bibinfo {author} {\bibfnamefont {D.}~\bibnamefont {Bourgund}}, \bibinfo
  {author} {\bibfnamefont {P.}~\bibnamefont {Bojovi\'c}}, \bibinfo {author}
  {\bibfnamefont {A.}~\bibnamefont {Bohrdt}}, \bibinfo {author} {\bibfnamefont
  {E.}~\bibnamefont {Demler}}, \bibinfo {author} {\bibfnamefont
  {F.}~\bibnamefont {Grusdt}}, \bibinfo {author} {\bibfnamefont
  {I.}~\bibnamefont {Bloch}},\ and\ \bibinfo {author} {\bibfnamefont
  {J.}~\bibnamefont {Zeiher}},\ }\bibfield  {title} {\bibinfo {title}
  {Magnetically mediated hole pairing in fermionic ladders of ultracold
  atoms},\ }\href {https://doi.org/10.1038/s41586-022-05437-y} {\bibfield
  {journal} {\bibinfo  {journal} {Nature}\ }\textbf {\bibinfo {volume} {613}},\
  \bibinfo {pages} {463} (\bibinfo {year} {2023})}\BibitemShut {NoStop}%
\bibitem [{\citenamefont {Wenz}\ \emph {et~al.}(2013)\citenamefont {Wenz},
  \citenamefont {Zürn}, \citenamefont {Murmann}, \citenamefont {Brouzos},
  \citenamefont {Lompe},\ and\ \citenamefont {Jochim}}]{Wenz2013}%
  \BibitemOpen
  \bibfield  {author} {\bibinfo {author} {\bibfnamefont {A.~N.}\ \bibnamefont
  {Wenz}}, \bibinfo {author} {\bibfnamefont {G.}~\bibnamefont {Zürn}},
  \bibinfo {author} {\bibfnamefont {S.}~\bibnamefont {Murmann}}, \bibinfo
  {author} {\bibfnamefont {I.}~\bibnamefont {Brouzos}}, \bibinfo {author}
  {\bibfnamefont {T.}~\bibnamefont {Lompe}},\ and\ \bibinfo {author}
  {\bibfnamefont {S.}~\bibnamefont {Jochim}},\ }\bibfield  {title} {\bibinfo
  {title} {From few to many: Observing the formation of a fermi sea one atom at
  a time},\ }\href {https://doi.org/10.1126/science.1240516} {\bibfield
  {journal} {\bibinfo  {journal} {Science}\ }\textbf {\bibinfo {volume}
  {342}},\ \bibinfo {pages} {457} (\bibinfo {year} {2013})}\BibitemShut
  {NoStop}%
\bibitem [{\citenamefont {Lee}\ \emph {et~al.}(2006)\citenamefont {Lee},
  \citenamefont {Nagaosa},\ and\ \citenamefont {Wen}}]{Lee2006}%
  \BibitemOpen
  \bibfield  {author} {\bibinfo {author} {\bibfnamefont {P.~A.}\ \bibnamefont
  {Lee}}, \bibinfo {author} {\bibfnamefont {N.}~\bibnamefont {Nagaosa}},\ and\
  \bibinfo {author} {\bibfnamefont {X.-G.}\ \bibnamefont {Wen}},\ }\bibfield
  {title} {\bibinfo {title} {Doping a mott insulator: Physics of
  high-temperature superconductivity},\ }\href
  {https://doi.org/10.1103/RevModPhys.78.17} {\bibfield  {journal} {\bibinfo
  {journal} {Rev. Mod. Phys.}\ }\textbf {\bibinfo {volume} {78}},\ \bibinfo
  {pages} {17} (\bibinfo {year} {2006})}\BibitemShut {NoStop}%
\bibitem [{\citenamefont {Xu}\ \emph {et~al.}(2022)\citenamefont {Xu},
  \citenamefont {Kendrick}, \citenamefont {Kale}, \citenamefont {Gang},
  \citenamefont {Ji}, \citenamefont {Scalettar}, \citenamefont {Lebrat},\ and\
  \citenamefont {Greiner}}]{greiner2023}%
  \BibitemOpen
  \bibfield  {author} {\bibinfo {author} {\bibfnamefont {M.}~\bibnamefont
  {Xu}}, \bibinfo {author} {\bibfnamefont {L.}~\bibnamefont {Kendrick}},
  \bibinfo {author} {\bibfnamefont {A.}~\bibnamefont {Kale}}, \bibinfo {author}
  {\bibfnamefont {Y.}~\bibnamefont {Gang}}, \bibinfo {author} {\bibfnamefont
  {G.}~\bibnamefont {Ji}}, \bibinfo {author} {\bibfnamefont {R.}~\bibnamefont
  {Scalettar}}, \bibinfo {author} {\bibfnamefont {M.}~\bibnamefont {Lebrat}},\
  and\ \bibinfo {author} {\bibfnamefont {M.}~\bibnamefont {Greiner}},\
  }\bibfield  {title} {\bibinfo {title} {Doping a frustrated fermi-hubbard
  magnet},\ }\bibfield  {journal} {\bibinfo  {journal} {arXiv:2212.13983}\
  }\href {https://doi.org/10.48550/arXiv.2212.13983}
  {10.48550/arXiv.2212.13983} (\bibinfo {year} {2022})\BibitemShut {NoStop}%
\bibitem [{\citenamefont {Anderson}\ \emph {et~al.}(2019)\citenamefont
  {Anderson}, \citenamefont {Wang}, \citenamefont {Xu}, \citenamefont {Venu},
  \citenamefont {Trotzky}, \citenamefont {Chevy},\ and\ \citenamefont
  {Thywissen}}]{Anderson2019}%
  \BibitemOpen
  \bibfield  {author} {\bibinfo {author} {\bibfnamefont {R.}~\bibnamefont
  {Anderson}}, \bibinfo {author} {\bibfnamefont {F.}~\bibnamefont {Wang}},
  \bibinfo {author} {\bibfnamefont {P.}~\bibnamefont {Xu}}, \bibinfo {author}
  {\bibfnamefont {V.}~\bibnamefont {Venu}}, \bibinfo {author} {\bibfnamefont
  {S.}~\bibnamefont {Trotzky}}, \bibinfo {author} {\bibfnamefont
  {F.}~\bibnamefont {Chevy}},\ and\ \bibinfo {author} {\bibfnamefont {J.~H.}\
  \bibnamefont {Thywissen}},\ }\bibfield  {title} {\bibinfo {title}
  {Conductivity spectrum of ultracold atoms in an optical lattice},\ }\href
  {https://doi.org/10.1103/PhysRevLett.122.153602} {\bibfield  {journal}
  {\bibinfo  {journal} {Phys. Rev. Lett.}\ }\textbf {\bibinfo {volume} {122}},\
  \bibinfo {pages} {153602} (\bibinfo {year} {2019})}\BibitemShut {NoStop}%
\bibitem [{\citenamefont {Brown}\ \emph {et~al.}(2019)\citenamefont {Brown},
  \citenamefont {Mitra}, \citenamefont {Guardado-Sanchez}, \citenamefont
  {Nourafkan}, \citenamefont {Reymbaut}, \citenamefont {H\'ebert},
  \citenamefont {Bergeron}, \citenamefont {Tremblay}, \citenamefont {Kokalj},
  \citenamefont {Huse}, \citenamefont {Schau{\ss}},\ and\ \citenamefont
  {Bakr}}]{Brown2019}%
  \BibitemOpen
  \bibfield  {author} {\bibinfo {author} {\bibfnamefont {P.~T.}\ \bibnamefont
  {Brown}}, \bibinfo {author} {\bibfnamefont {D.}~\bibnamefont {Mitra}},
  \bibinfo {author} {\bibfnamefont {E.}~\bibnamefont {Guardado-Sanchez}},
  \bibinfo {author} {\bibfnamefont {R.}~\bibnamefont {Nourafkan}}, \bibinfo
  {author} {\bibfnamefont {A.}~\bibnamefont {Reymbaut}}, \bibinfo {author}
  {\bibfnamefont {C.-D.}\ \bibnamefont {H\'ebert}}, \bibinfo {author}
  {\bibfnamefont {S.}~\bibnamefont {Bergeron}}, \bibinfo {author}
  {\bibfnamefont {A.-M.~S.}\ \bibnamefont {Tremblay}}, \bibinfo {author}
  {\bibfnamefont {J.}~\bibnamefont {Kokalj}}, \bibinfo {author} {\bibfnamefont
  {D.~A.}\ \bibnamefont {Huse}}, \bibinfo {author} {\bibfnamefont
  {P.}~\bibnamefont {Schau{\ss}}},\ and\ \bibinfo {author} {\bibfnamefont
  {W.~S.}\ \bibnamefont {Bakr}},\ }\bibfield  {title} {\bibinfo {title} {Bad
  metallic transport in a cold atom fermi-hubbard system},\ }\href
  {https://doi.org/10.1126/science.aat4134} {\bibfield  {journal} {\bibinfo
  {journal} {Science}\ }\textbf {\bibinfo {volume} {363}},\ \bibinfo {pages}
  {379} (\bibinfo {year} {2019})}\BibitemShut {NoStop}%
\bibitem [{\citenamefont {Nichols}\ \emph {et~al.}(2019)\citenamefont
  {Nichols}, \citenamefont {Cheuk}, \citenamefont {Okan}, \citenamefont
  {Hartke}, \citenamefont {Mendez}, \citenamefont {Senthil}, \citenamefont
  {Khatami}, \citenamefont {Zhang},\ and\ \citenamefont
  {Zwierlein}}]{Nichols2019}%
  \BibitemOpen
  \bibfield  {author} {\bibinfo {author} {\bibfnamefont {M.~A.}\ \bibnamefont
  {Nichols}}, \bibinfo {author} {\bibfnamefont {L.~W.}\ \bibnamefont {Cheuk}},
  \bibinfo {author} {\bibfnamefont {M.}~\bibnamefont {Okan}}, \bibinfo {author}
  {\bibfnamefont {T.~R.}\ \bibnamefont {Hartke}}, \bibinfo {author}
  {\bibfnamefont {E.}~\bibnamefont {Mendez}}, \bibinfo {author} {\bibfnamefont
  {T.}~\bibnamefont {Senthil}}, \bibinfo {author} {\bibfnamefont
  {E.}~\bibnamefont {Khatami}}, \bibinfo {author} {\bibfnamefont
  {H.}~\bibnamefont {Zhang}},\ and\ \bibinfo {author} {\bibfnamefont {M.~W.}\
  \bibnamefont {Zwierlein}},\ }\bibfield  {title} {\bibinfo {title} {Spin
  transport in a mott insulator of ultracold fermions},\ }\href
  {https://doi.org/10.1126/science.aat4387} {\bibfield  {journal} {\bibinfo
  {journal} {Science}\ }\textbf {\bibinfo {volume} {363}},\ \bibinfo {pages}
  {383} (\bibinfo {year} {2019})}\BibitemShut {NoStop}%
\bibitem [{\citenamefont {Chiu}\ \emph {et~al.}(2019)\citenamefont {Chiu},
  \citenamefont {Ji}, \citenamefont {Bohrdt}, \citenamefont {Xu}, \citenamefont
  {Knap}, \citenamefont {Demler}, \citenamefont {Grusdt}, \citenamefont
  {Greiner},\ and\ \citenamefont {Greif}}]{Chiu2019}%
  \BibitemOpen
  \bibfield  {author} {\bibinfo {author} {\bibfnamefont {C.~S.}\ \bibnamefont
  {Chiu}}, \bibinfo {author} {\bibfnamefont {G.}~\bibnamefont {Ji}}, \bibinfo
  {author} {\bibfnamefont {A.}~\bibnamefont {Bohrdt}}, \bibinfo {author}
  {\bibfnamefont {M.}~\bibnamefont {Xu}}, \bibinfo {author} {\bibfnamefont
  {M.}~\bibnamefont {Knap}}, \bibinfo {author} {\bibfnamefont {E.}~\bibnamefont
  {Demler}}, \bibinfo {author} {\bibfnamefont {F.}~\bibnamefont {Grusdt}},
  \bibinfo {author} {\bibfnamefont {M.}~\bibnamefont {Greiner}},\ and\ \bibinfo
  {author} {\bibfnamefont {D.}~\bibnamefont {Greif}},\ }\bibfield  {title}
  {\bibinfo {title} {{String patterns in the doped Hubbard model}},\ }\href
  {https://doi.org/10.1126/science.aav3587} {\bibfield  {journal} {\bibinfo
  {journal} {Science}\ }\textbf {\bibinfo {volume} {365}},\ \bibinfo {pages}
  {251} (\bibinfo {year} {2019})}\BibitemShut {NoStop}%
\bibitem [{\citenamefont {Salomon}\ \emph {et~al.}(2019)\citenamefont
  {Salomon}, \citenamefont {Koepsell}, \citenamefont {Vijayan}, \citenamefont
  {Hilker}, \citenamefont {Nespolo}, \citenamefont {Pollet}, \citenamefont
  {Bloch},\ and\ \citenamefont {Gross}}]{Salomon2019}%
  \BibitemOpen
  \bibfield  {author} {\bibinfo {author} {\bibfnamefont {G.}~\bibnamefont
  {Salomon}}, \bibinfo {author} {\bibfnamefont {J.}~\bibnamefont {Koepsell}},
  \bibinfo {author} {\bibfnamefont {J.}~\bibnamefont {Vijayan}}, \bibinfo
  {author} {\bibfnamefont {T.~A.}\ \bibnamefont {Hilker}}, \bibinfo {author}
  {\bibfnamefont {J.}~\bibnamefont {Nespolo}}, \bibinfo {author} {\bibfnamefont
  {L.}~\bibnamefont {Pollet}}, \bibinfo {author} {\bibfnamefont
  {I.}~\bibnamefont {Bloch}},\ and\ \bibinfo {author} {\bibfnamefont
  {C.}~\bibnamefont {Gross}},\ }\bibfield  {title} {\bibinfo {title} {Direct
  observation of incommensurate magnetism in hubbard chains},\ }\href
  {https://doi.org/10.1038/s41586-018-0778-7} {\bibfield  {journal} {\bibinfo
  {journal} {Nature}\ }\textbf {\bibinfo {volume} {565}},\ \bibinfo {pages}
  {56} (\bibinfo {year} {2019})}\BibitemShut {NoStop}%
\bibitem [{\citenamefont {Koepsell}\ \emph {et~al.}(2019)\citenamefont
  {Koepsell}, \citenamefont {Vijayan}, \citenamefont {Sompet}, \citenamefont
  {Grusdt}, \citenamefont {Hilker}, \citenamefont {Demler}, \citenamefont
  {Salomon}, \citenamefont {Bloch},\ and\ \citenamefont
  {Gross}}]{Koepsell2019}%
  \BibitemOpen
  \bibfield  {author} {\bibinfo {author} {\bibfnamefont {J.}~\bibnamefont
  {Koepsell}}, \bibinfo {author} {\bibfnamefont {J.}~\bibnamefont {Vijayan}},
  \bibinfo {author} {\bibfnamefont {P.}~\bibnamefont {Sompet}}, \bibinfo
  {author} {\bibfnamefont {F.}~\bibnamefont {Grusdt}}, \bibinfo {author}
  {\bibfnamefont {T.~A.}\ \bibnamefont {Hilker}}, \bibinfo {author}
  {\bibfnamefont {E.}~\bibnamefont {Demler}}, \bibinfo {author} {\bibfnamefont
  {G.}~\bibnamefont {Salomon}}, \bibinfo {author} {\bibfnamefont
  {I.}~\bibnamefont {Bloch}},\ and\ \bibinfo {author} {\bibfnamefont
  {C.}~\bibnamefont {Gross}},\ }\bibfield  {title} {\bibinfo {title} {Imaging
  magnetic polarons in the doped fermi–hubbard model},\ }\href
  {https://doi.org/10.1038/s41586-019-1463-1} {\bibfield  {journal} {\bibinfo
  {journal} {Nature}\ }\textbf {\bibinfo {volume} {572}},\ \bibinfo {pages}
  {358} (\bibinfo {year} {2019})}\BibitemShut {NoStop}%
\bibitem [{\citenamefont {Koepsell}\ \emph {et~al.}(2021)\citenamefont
  {Koepsell}, \citenamefont {Bourgund}, \citenamefont {Sompet}, \citenamefont
  {Hirthe}, \citenamefont {Bohrdt}, \citenamefont {Wang}, \citenamefont
  {Grusdt}, \citenamefont {Demler}, \citenamefont {Salomon}, \citenamefont
  {Gross},\ and\ \citenamefont {Bloch}}]{Koepsell2021}%
  \BibitemOpen
  \bibfield  {author} {\bibinfo {author} {\bibfnamefont {J.}~\bibnamefont
  {Koepsell}}, \bibinfo {author} {\bibfnamefont {D.}~\bibnamefont {Bourgund}},
  \bibinfo {author} {\bibfnamefont {P.}~\bibnamefont {Sompet}}, \bibinfo
  {author} {\bibfnamefont {S.}~\bibnamefont {Hirthe}}, \bibinfo {author}
  {\bibfnamefont {A.}~\bibnamefont {Bohrdt}}, \bibinfo {author} {\bibfnamefont
  {Y.}~\bibnamefont {Wang}}, \bibinfo {author} {\bibfnamefont {F.}~\bibnamefont
  {Grusdt}}, \bibinfo {author} {\bibfnamefont {E.}~\bibnamefont {Demler}},
  \bibinfo {author} {\bibfnamefont {G.}~\bibnamefont {Salomon}}, \bibinfo
  {author} {\bibfnamefont {C.}~\bibnamefont {Gross}},\ and\ \bibinfo {author}
  {\bibfnamefont {I.}~\bibnamefont {Bloch}},\ }\bibfield  {title} {\bibinfo
  {title} {Microscopic evolution of doped mott insulators from polaronic metal
  to fermi liquid},\ }\href {https://doi.org/10.1126/science.abe7165}
  {\bibfield  {journal} {\bibinfo  {journal} {Science}\ }\textbf {\bibinfo
  {volume} {374}},\ \bibinfo {pages} {82} (\bibinfo {year} {2021})}\BibitemShut
  {NoStop}%
\bibitem [{\citenamefont {Ji}\ \emph {et~al.}(2021)\citenamefont {Ji},
  \citenamefont {Xu}, \citenamefont {Kendrick}, \citenamefont {Chiu},
  \citenamefont {Br{\"{u}}ggenj{\"{u}}rgen}, \citenamefont {Greif},
  \citenamefont {Bohrdt}, \citenamefont {Grusdt}, \citenamefont {Knap},
  \citenamefont {Demler}, \citenamefont {Lebrat},\ and\ \citenamefont
  {Greiner}}]{Ji2021}%
  \BibitemOpen
  \bibfield  {author} {\bibinfo {author} {\bibfnamefont {G.}~\bibnamefont
  {Ji}}, \bibinfo {author} {\bibfnamefont {M.}~\bibnamefont {Xu}}, \bibinfo
  {author} {\bibfnamefont {L.~H.}\ \bibnamefont {Kendrick}}, \bibinfo {author}
  {\bibfnamefont {C.~S.}\ \bibnamefont {Chiu}}, \bibinfo {author}
  {\bibfnamefont {J.~C.}\ \bibnamefont {Br{\"{u}}ggenj{\"{u}}rgen}}, \bibinfo
  {author} {\bibfnamefont {D.}~\bibnamefont {Greif}}, \bibinfo {author}
  {\bibfnamefont {A.}~\bibnamefont {Bohrdt}}, \bibinfo {author} {\bibfnamefont
  {F.}~\bibnamefont {Grusdt}}, \bibinfo {author} {\bibfnamefont
  {M.}~\bibnamefont {Knap}}, \bibinfo {author} {\bibfnamefont {E.}~\bibnamefont
  {Demler}}, \bibinfo {author} {\bibfnamefont {M.}~\bibnamefont {Lebrat}},\
  and\ \bibinfo {author} {\bibfnamefont {M.}~\bibnamefont {Greiner}},\
  }\bibfield  {title} {\bibinfo {title} {{Coupling a Mobile Hole to an
  Antiferromagnetic Spin Background: Transient Dynamics of a Magnetic
  Polaron}},\ }\href {https://doi.org/10.1103/PhysRevX.11.021022} {\bibfield
  {journal} {\bibinfo  {journal} {Phys. Rev. X}\ }\textbf {\bibinfo {volume}
  {11}},\ \bibinfo {pages} {021022} (\bibinfo {year} {2021})}\BibitemShut
  {NoStop}%
\bibitem [{\citenamefont {Fisher}\ \emph {et~al.}(1989)\citenamefont {Fisher},
  \citenamefont {Weichman}, \citenamefont {Grinstein},\ and\ \citenamefont
  {Fisher}}]{Fisher1989}%
  \BibitemOpen
  \bibfield  {author} {\bibinfo {author} {\bibfnamefont {M.~P.~A.}\
  \bibnamefont {Fisher}}, \bibinfo {author} {\bibfnamefont {P.~B.}\
  \bibnamefont {Weichman}}, \bibinfo {author} {\bibfnamefont {G.}~\bibnamefont
  {Grinstein}},\ and\ \bibinfo {author} {\bibfnamefont {D.~S.}\ \bibnamefont
  {Fisher}},\ }\bibfield  {title} {\bibinfo {title} {Boson localization and the
  superfluid-insulator transition},\ }\href
  {https://doi.org/10.1103/PhysRevB.40.546} {\bibfield  {journal} {\bibinfo
  {journal} {Phys. Rev. B}\ }\textbf {\bibinfo {volume} {40}},\ \bibinfo
  {pages} {546} (\bibinfo {year} {1989})}\BibitemShut {NoStop}%
\bibitem [{\citenamefont {Jaksch}\ \emph {et~al.}(1998)\citenamefont {Jaksch},
  \citenamefont {Bruder}, \citenamefont {Cirac}, \citenamefont {Gardiner},\
  and\ \citenamefont {Zoller}}]{Jaksch1998}%
  \BibitemOpen
  \bibfield  {author} {\bibinfo {author} {\bibfnamefont {D.}~\bibnamefont
  {Jaksch}}, \bibinfo {author} {\bibfnamefont {C.}~\bibnamefont {Bruder}},
  \bibinfo {author} {\bibfnamefont {J.~I.}\ \bibnamefont {Cirac}}, \bibinfo
  {author} {\bibfnamefont {C.~W.}\ \bibnamefont {Gardiner}},\ and\ \bibinfo
  {author} {\bibfnamefont {P.}~\bibnamefont {Zoller}},\ }\bibfield  {title}
  {\bibinfo {title} {Cold bosonic atoms in optical lattices},\ }\href
  {https://doi.org/10.1103/PhysRevLett.81.3108} {\bibfield  {journal} {\bibinfo
   {journal} {Phys. Rev. Lett.}\ }\textbf {\bibinfo {volume} {81}},\ \bibinfo
  {pages} {3108} (\bibinfo {year} {1998})}\BibitemShut {NoStop}%
\bibitem [{\citenamefont {Lazarides}\ \emph {et~al.}(2011)\citenamefont
  {Lazarides}, \citenamefont {Tieleman},\ and\ \citenamefont {{Morais
  Smith}}}]{Lazarides2011}%
  \BibitemOpen
  \bibfield  {author} {\bibinfo {author} {\bibfnamefont {A.}~\bibnamefont
  {Lazarides}}, \bibinfo {author} {\bibfnamefont {O.}~\bibnamefont
  {Tieleman}},\ and\ \bibinfo {author} {\bibfnamefont {C.}~\bibnamefont
  {{Morais Smith}}},\ }\bibfield  {title} {\bibinfo {title} {{Strongly
  interacting bosons in a one-dimensional optical lattice at incommensurate
  densities}},\ }\href {https://doi.org/10.1103/PhysRevA.84.023620} {\bibfield
  {journal} {\bibinfo  {journal} {Phys. Rev. A}\ }\textbf {\bibinfo {volume}
  {84}},\ \bibinfo {pages} {023620} (\bibinfo {year} {2011})}\BibitemShut
  {NoStop}%
\bibitem [{\citenamefont {B{\"{u}}chler}(2011)}]{Buchler2011}%
  \BibitemOpen
  \bibfield  {author} {\bibinfo {author} {\bibfnamefont {H.~P.}\ \bibnamefont
  {B{\"{u}}chler}},\ }\bibfield  {title} {\bibinfo {title} {{Crystalline phase
  for one-dimensional ultra-cold atomic bosons}},\ }\href
  {https://doi.org/10.1088/1367-2630/13/9/093040} {\bibfield  {journal}
  {\bibinfo  {journal} {New J. Phys.}\ }\textbf {\bibinfo {volume} {13}},\
  \bibinfo {pages} {093040} (\bibinfo {year} {2011})}\BibitemShut {NoStop}%
\bibitem [{\citenamefont {Damski}\ \emph {et~al.}(2003)\citenamefont {Damski},
  \citenamefont {Zakrzewski}, \citenamefont {Santos}, \citenamefont {Zoller},\
  and\ \citenamefont {Lewenstein}}]{Damski2003}%
  \BibitemOpen
  \bibfield  {author} {\bibinfo {author} {\bibfnamefont {B.}~\bibnamefont
  {Damski}}, \bibinfo {author} {\bibfnamefont {J.}~\bibnamefont {Zakrzewski}},
  \bibinfo {author} {\bibfnamefont {L.}~\bibnamefont {Santos}}, \bibinfo
  {author} {\bibfnamefont {P.}~\bibnamefont {Zoller}},\ and\ \bibinfo {author}
  {\bibfnamefont {M.}~\bibnamefont {Lewenstein}},\ }\bibfield  {title}
  {\bibinfo {title} {Atomic bose and anderson glasses in optical lattices},\
  }\href {https://doi.org/10.1103/PhysRevLett.91.080403} {\bibfield  {journal}
  {\bibinfo  {journal} {Phys. Rev. Lett.}\ }\textbf {\bibinfo {volume} {91}},\
  \bibinfo {pages} {080403} (\bibinfo {year} {2003})}\BibitemShut {NoStop}%
\bibitem [{\citenamefont {Fallani}\ \emph {et~al.}(2007)\citenamefont
  {Fallani}, \citenamefont {Lye}, \citenamefont {Guarrera}, \citenamefont
  {Fort},\ and\ \citenamefont {Inguscio}}]{Fallani2007}%
  \BibitemOpen
  \bibfield  {author} {\bibinfo {author} {\bibfnamefont {L.}~\bibnamefont
  {Fallani}}, \bibinfo {author} {\bibfnamefont {J.~E.}\ \bibnamefont {Lye}},
  \bibinfo {author} {\bibfnamefont {V.}~\bibnamefont {Guarrera}}, \bibinfo
  {author} {\bibfnamefont {C.}~\bibnamefont {Fort}},\ and\ \bibinfo {author}
  {\bibfnamefont {M.}~\bibnamefont {Inguscio}},\ }\bibfield  {title} {\bibinfo
  {title} {Ultracold atoms in a disordered crystal of light: Towards a bose
  glass},\ }\href {https://doi.org/10.1103/PhysRevLett.98.130404} {\bibfield
  {journal} {\bibinfo  {journal} {Phys. Rev. Lett.}\ }\textbf {\bibinfo
  {volume} {98}},\ \bibinfo {pages} {130404} (\bibinfo {year}
  {2007})}\BibitemShut {NoStop}%
\bibitem [{\citenamefont {Cai}\ \emph {et~al.}(2010)\citenamefont {Cai},
  \citenamefont {Chen},\ and\ \citenamefont {Wang}}]{Cai2010}%
  \BibitemOpen
  \bibfield  {author} {\bibinfo {author} {\bibfnamefont {X.}~\bibnamefont
  {Cai}}, \bibinfo {author} {\bibfnamefont {S.}~\bibnamefont {Chen}},\ and\
  \bibinfo {author} {\bibfnamefont {Y.}~\bibnamefont {Wang}},\ }\bibfield
  {title} {\bibinfo {title} {Superfluid-to-bose-glass transition of hard-core
  bosons in a one-dimensional incommensurate optical lattice},\ }\href
  {https://doi.org/10.1103/PhysRevA.81.023626} {\bibfield  {journal} {\bibinfo
  {journal} {Phys. Rev. A}\ }\textbf {\bibinfo {volume} {81}},\ \bibinfo
  {pages} {023626} (\bibinfo {year} {2010})}\BibitemShut {NoStop}%
\bibitem [{\citenamefont {Astrakharchik}\ \emph {et~al.}(2017)\citenamefont
  {Astrakharchik}, \citenamefont {Krutitsky}, \citenamefont {Lewenstein},
  \citenamefont {Mazzanti},\ and\ \citenamefont {Boronat}}]{Astrakharchik2017}%
  \BibitemOpen
  \bibfield  {author} {\bibinfo {author} {\bibfnamefont {G.~E.}\ \bibnamefont
  {Astrakharchik}}, \bibinfo {author} {\bibfnamefont {K.~V.}\ \bibnamefont
  {Krutitsky}}, \bibinfo {author} {\bibfnamefont {M.}~\bibnamefont
  {Lewenstein}}, \bibinfo {author} {\bibfnamefont {F.}~\bibnamefont
  {Mazzanti}},\ and\ \bibinfo {author} {\bibfnamefont {J.}~\bibnamefont
  {Boronat}},\ }\bibfield  {title} {\bibinfo {title} {Optical lattices as a
  tool to study defect-induced superfluidity},\ }\href
  {https://doi.org/10.1103/PhysRevA.96.033606} {\bibfield  {journal} {\bibinfo
  {journal} {Phys. Rev. A}\ }\textbf {\bibinfo {volume} {96}},\ \bibinfo
  {pages} {033606} (\bibinfo {year} {2017})}\BibitemShut {NoStop}%
\bibitem [{\citenamefont {Kwon}\ \emph {et~al.}(2021)\citenamefont {Kwon},
  \citenamefont {Del~Pace}, \citenamefont {Xhani}, \citenamefont {Galantucci},
  \citenamefont {Muzi~Falconi}, \citenamefont {Inguscio}, \citenamefont
  {Scazza},\ and\ \citenamefont {Roati}}]{Kwon2021}%
  \BibitemOpen
  \bibfield  {author} {\bibinfo {author} {\bibfnamefont {W.~J.}\ \bibnamefont
  {Kwon}}, \bibinfo {author} {\bibfnamefont {G.}~\bibnamefont {Del~Pace}},
  \bibinfo {author} {\bibfnamefont {K.}~\bibnamefont {Xhani}}, \bibinfo
  {author} {\bibfnamefont {L.}~\bibnamefont {Galantucci}}, \bibinfo {author}
  {\bibfnamefont {A.}~\bibnamefont {Muzi~Falconi}}, \bibinfo {author}
  {\bibfnamefont {M.}~\bibnamefont {Inguscio}}, \bibinfo {author}
  {\bibfnamefont {F.}~\bibnamefont {Scazza}},\ and\ \bibinfo {author}
  {\bibfnamefont {G.}~\bibnamefont {Roati}},\ }\bibfield  {title} {\bibinfo
  {title} {Sound emission and annihilations in a programmable quantum vortex
  collider},\ }\href {https://doi.org/10.1038/s41586-021-04047-4} {\bibfield
  {journal} {\bibinfo  {journal} {Nature}\ }\textbf {\bibinfo {volume} {600}},\
  \bibinfo {pages} {64} (\bibinfo {year} {2021})}\BibitemShut {NoStop}%
\bibitem [{\citenamefont {Del~Pace}\ \emph {et~al.}(2022)\citenamefont
  {Del~Pace}, \citenamefont {Xhani}, \citenamefont {Muzi~Falconi},
  \citenamefont {Fedrizzi}, \citenamefont {Grani}, \citenamefont
  {Hernandez~Rajkov}, \citenamefont {Inguscio}, \citenamefont {Scazza},
  \citenamefont {Kwon},\ and\ \citenamefont {Roati}}]{Pace2022}%
  \BibitemOpen
  \bibfield  {author} {\bibinfo {author} {\bibfnamefont {G.}~\bibnamefont
  {Del~Pace}}, \bibinfo {author} {\bibfnamefont {K.}~\bibnamefont {Xhani}},
  \bibinfo {author} {\bibfnamefont {A.}~\bibnamefont {Muzi~Falconi}}, \bibinfo
  {author} {\bibfnamefont {M.}~\bibnamefont {Fedrizzi}}, \bibinfo {author}
  {\bibfnamefont {N.}~\bibnamefont {Grani}}, \bibinfo {author} {\bibfnamefont
  {D.}~\bibnamefont {Hernandez~Rajkov}}, \bibinfo {author} {\bibfnamefont
  {M.}~\bibnamefont {Inguscio}}, \bibinfo {author} {\bibfnamefont
  {F.}~\bibnamefont {Scazza}}, \bibinfo {author} {\bibfnamefont {W.~J.}\
  \bibnamefont {Kwon}},\ and\ \bibinfo {author} {\bibfnamefont
  {G.}~\bibnamefont {Roati}},\ }\bibfield  {title} {\bibinfo {title}
  {Imprinting persistent currents in tunable fermionic rings},\ }\href
  {https://doi.org/10.1103/PhysRevX.12.041037} {\bibfield  {journal} {\bibinfo
  {journal} {Phys. Rev. X}\ }\textbf {\bibinfo {volume} {12}},\ \bibinfo
  {pages} {041037} (\bibinfo {year} {2022})}\BibitemShut {NoStop}%
\bibitem [{\citenamefont {Reeves}\ \emph {et~al.}(2022)\citenamefont {Reeves},
  \citenamefont {Goddard-Lee}, \citenamefont {Gauthier}, \citenamefont
  {Stockdale}, \citenamefont {Salman}, \citenamefont {Edmonds}, \citenamefont
  {Yu}, \citenamefont {Bradley}, \citenamefont {Baker}, \citenamefont
  {Rubinsztein-Dunlop}, \citenamefont {Davis},\ and\ \citenamefont
  {Neely}}]{Reeves2022}%
  \BibitemOpen
  \bibfield  {author} {\bibinfo {author} {\bibfnamefont {M.~T.}\ \bibnamefont
  {Reeves}}, \bibinfo {author} {\bibfnamefont {K.}~\bibnamefont {Goddard-Lee}},
  \bibinfo {author} {\bibfnamefont {G.}~\bibnamefont {Gauthier}}, \bibinfo
  {author} {\bibfnamefont {O.~R.}\ \bibnamefont {Stockdale}}, \bibinfo {author}
  {\bibfnamefont {H.}~\bibnamefont {Salman}}, \bibinfo {author} {\bibfnamefont
  {T.}~\bibnamefont {Edmonds}}, \bibinfo {author} {\bibfnamefont
  {X.}~\bibnamefont {Yu}}, \bibinfo {author} {\bibfnamefont {A.~S.}\
  \bibnamefont {Bradley}}, \bibinfo {author} {\bibfnamefont {M.}~\bibnamefont
  {Baker}}, \bibinfo {author} {\bibfnamefont {H.}~\bibnamefont
  {Rubinsztein-Dunlop}}, \bibinfo {author} {\bibfnamefont {M.~J.}\ \bibnamefont
  {Davis}},\ and\ \bibinfo {author} {\bibfnamefont {T.~W.}\ \bibnamefont
  {Neely}},\ }\bibfield  {title} {\bibinfo {title} {Turbulent relaxation to
  equilibrium in a two-dimensional quantum vortex gas},\ }\href
  {https://doi.org/10.1103/PhysRevX.12.011031} {\bibfield  {journal} {\bibinfo
  {journal} {Phys. Rev. X}\ }\textbf {\bibinfo {volume} {12}},\ \bibinfo
  {pages} {011031} (\bibinfo {year} {2022})}\BibitemShut {NoStop}%
\bibitem [{\citenamefont {Brouzos}\ \emph {et~al.}(2010)\citenamefont
  {Brouzos}, \citenamefont {Z{\"{o}}llner},\ and\ \citenamefont
  {Schmelcher}}]{Brouzos2010}%
  \BibitemOpen
  \bibfield  {author} {\bibinfo {author} {\bibfnamefont {I.}~\bibnamefont
  {Brouzos}}, \bibinfo {author} {\bibfnamefont {S.}~\bibnamefont
  {Z{\"{o}}llner}},\ and\ \bibinfo {author} {\bibfnamefont {P.}~\bibnamefont
  {Schmelcher}},\ }\bibfield  {title} {\bibinfo {title} {{Correlation versus
  commensurability effects for finite bosonic systems in one-dimensional
  lattices}},\ }\href {https://doi.org/10.1103/PhysRevA.81.053613} {\bibfield
  {journal} {\bibinfo  {journal} {Phys. Rev. A}\ }\textbf {\bibinfo {volume}
  {81}},\ \bibinfo {pages} {053613} (\bibinfo {year} {2010})}\BibitemShut
  {NoStop}%
\bibitem [{\citenamefont {Yamamoto}\ \emph {et~al.}(2020)\citenamefont
  {Yamamoto}, \citenamefont {Ozawa}, \citenamefont {Nak}, \citenamefont
  {Nakamura},\ and\ \citenamefont {Fukuhara}}]{Yamamoto2020}%
  \BibitemOpen
  \bibfield  {author} {\bibinfo {author} {\bibfnamefont {R.}~\bibnamefont
  {Yamamoto}}, \bibinfo {author} {\bibfnamefont {H.}~\bibnamefont {Ozawa}},
  \bibinfo {author} {\bibfnamefont {D.~C.}\ \bibnamefont {Nak}}, \bibinfo
  {author} {\bibfnamefont {I.}~\bibnamefont {Nakamura}},\ and\ \bibinfo
  {author} {\bibfnamefont {T.}~\bibnamefont {Fukuhara}},\ }\bibfield  {title}
  {\bibinfo {title} {Single-site-resolved imaging of ultracold atoms in a
  triangular optical lattice},\ }\href
  {https://doi.org/10.1088/1367-2630/abcdc8} {\bibfield  {journal} {\bibinfo
  {journal} {New J. Phys.}\ }\textbf {\bibinfo {volume} {22}},\ \bibinfo
  {pages} {123028} (\bibinfo {year} {2020})}\BibitemShut {NoStop}%
\bibitem [{\citenamefont {Yang}\ \emph {et~al.}(2021)\citenamefont {Yang},
  \citenamefont {Liu}, \citenamefont {Mongkolkiattichai},\ and\ \citenamefont
  {Schauss}}]{Yang2021}%
  \BibitemOpen
  \bibfield  {author} {\bibinfo {author} {\bibfnamefont {J.}~\bibnamefont
  {Yang}}, \bibinfo {author} {\bibfnamefont {L.}~\bibnamefont {Liu}}, \bibinfo
  {author} {\bibfnamefont {J.}~\bibnamefont {Mongkolkiattichai}},\ and\
  \bibinfo {author} {\bibfnamefont {P.}~\bibnamefont {Schauss}},\ }\bibfield
  {title} {\bibinfo {title} {Site-resolved imaging of ultracold fermions in a
  triangular-lattice quantum gas microscope},\ }\href
  {https://doi.org/10.1103/PRXQuantum.2.020344} {\bibfield  {journal} {\bibinfo
   {journal} {PRX Quantum}\ }\textbf {\bibinfo {volume} {2}},\ \bibinfo {pages}
  {020344} (\bibinfo {year} {2021})}\BibitemShut {NoStop}%
\bibitem [{\citenamefont {Fukuhara}\ \emph {et~al.}(2013)\citenamefont
  {Fukuhara}, \citenamefont {Kantian}, \citenamefont {Endres}, \citenamefont
  {Cheneau}, \citenamefont {Schau{\ss}}, \citenamefont {Hild}, \citenamefont
  {Bellem}, \citenamefont {Schollw{\"{o}}ck}, \citenamefont {Giamarchi},
  \citenamefont {Gross}, \citenamefont {Bloch},\ and\ \citenamefont
  {Kuhr}}]{Fukuhara2013a}%
  \BibitemOpen
  \bibfield  {author} {\bibinfo {author} {\bibfnamefont {T.}~\bibnamefont
  {Fukuhara}}, \bibinfo {author} {\bibfnamefont {A.}~\bibnamefont {Kantian}},
  \bibinfo {author} {\bibfnamefont {M.}~\bibnamefont {Endres}}, \bibinfo
  {author} {\bibfnamefont {M.}~\bibnamefont {Cheneau}}, \bibinfo {author}
  {\bibfnamefont {P.}~\bibnamefont {Schau{\ss}}}, \bibinfo {author}
  {\bibfnamefont {S.}~\bibnamefont {Hild}}, \bibinfo {author} {\bibfnamefont
  {D.}~\bibnamefont {Bellem}}, \bibinfo {author} {\bibfnamefont
  {U.}~\bibnamefont {Schollw{\"{o}}ck}}, \bibinfo {author} {\bibfnamefont
  {T.}~\bibnamefont {Giamarchi}}, \bibinfo {author} {\bibfnamefont
  {C.}~\bibnamefont {Gross}}, \bibinfo {author} {\bibfnamefont
  {I.}~\bibnamefont {Bloch}},\ and\ \bibinfo {author} {\bibfnamefont
  {S.}~\bibnamefont {Kuhr}},\ }\bibfield  {title} {\bibinfo {title} {{Quantum
  dynamics of a mobile spin impurity}},\ }\href
  {https://doi.org/10.1038/nphys2561} {\bibfield  {journal} {\bibinfo
  {journal} {Nat. Phys.}\ }\textbf {\bibinfo {volume} {9}},\ \bibinfo {pages}
  {235} (\bibinfo {year} {2013})}\BibitemShut {NoStop}%
\bibitem [{\citenamefont {Endres}\ \emph {et~al.}(2011)\citenamefont {Endres},
  \citenamefont {Cheneau}, \citenamefont {Fukuhara}, \citenamefont
  {Weitenberg}, \citenamefont {Schauss}, \citenamefont {Gross}, \citenamefont
  {Mazza}, \citenamefont {Banuls}, \citenamefont {Pollet}, \citenamefont
  {Bloch},\ and\ \citenamefont {Kuhr}}]{Endres2011}%
  \BibitemOpen
  \bibfield  {author} {\bibinfo {author} {\bibfnamefont {M.}~\bibnamefont
  {Endres}}, \bibinfo {author} {\bibfnamefont {M.}~\bibnamefont {Cheneau}},
  \bibinfo {author} {\bibfnamefont {T.}~\bibnamefont {Fukuhara}}, \bibinfo
  {author} {\bibfnamefont {C.}~\bibnamefont {Weitenberg}}, \bibinfo {author}
  {\bibfnamefont {P.}~\bibnamefont {Schauss}}, \bibinfo {author} {\bibfnamefont
  {C.}~\bibnamefont {Gross}}, \bibinfo {author} {\bibfnamefont
  {L.}~\bibnamefont {Mazza}}, \bibinfo {author} {\bibfnamefont {M.~C.}\
  \bibnamefont {Banuls}}, \bibinfo {author} {\bibfnamefont {L.}~\bibnamefont
  {Pollet}}, \bibinfo {author} {\bibfnamefont {I.}~\bibnamefont {Bloch}},\ and\
  \bibinfo {author} {\bibfnamefont {S.}~\bibnamefont {Kuhr}},\ }\bibfield
  {title} {\bibinfo {title} {Observation of correlated particle-hole pairs and
  string order in low-dimensional mott insulators},\ }\href
  {https://doi.org/10.1126/science.120928} {\bibfield  {journal} {\bibinfo
  {journal} {Science}\ }\textbf {\bibinfo {volume} {334}},\ \bibinfo {pages}
  {200} (\bibinfo {year} {2011})}\BibitemShut {NoStop}%
\bibitem [{\citenamefont {Rigol}\ \emph {et~al.}(2009)\citenamefont {Rigol},
  \citenamefont {Batrouni}, \citenamefont {Rousseau},\ and\ \citenamefont
  {Scalettar}}]{Rigol2009}%
  \BibitemOpen
  \bibfield  {author} {\bibinfo {author} {\bibfnamefont {M.}~\bibnamefont
  {Rigol}}, \bibinfo {author} {\bibfnamefont {G.~G.}\ \bibnamefont {Batrouni}},
  \bibinfo {author} {\bibfnamefont {V.~G.}\ \bibnamefont {Rousseau}},\ and\
  \bibinfo {author} {\bibfnamefont {R.~T.}\ \bibnamefont {Scalettar}},\
  }\bibfield  {title} {\bibinfo {title} {State diagrams for harmonically
  trapped bosons in optical lattices},\ }\href
  {https://doi.org/10.1103/PhysRevA.79.053605} {\bibfield  {journal} {\bibinfo
  {journal} {Phys. Rev. A}\ }\textbf {\bibinfo {volume} {79}},\ \bibinfo
  {pages} {053605} (\bibinfo {year} {2009})}\BibitemShut {NoStop}%
\bibitem [{\citenamefont {Gemelke}\ \emph {et~al.}(2009)\citenamefont
  {Gemelke}, \citenamefont {Zhang}, \citenamefont {Hung},\ and\ \citenamefont
  {Chin}}]{Gemelke2009}%
  \BibitemOpen
  \bibfield  {author} {\bibinfo {author} {\bibfnamefont {N.}~\bibnamefont
  {Gemelke}}, \bibinfo {author} {\bibfnamefont {X.}~\bibnamefont {Zhang}},
  \bibinfo {author} {\bibfnamefont {C.~L.}\ \bibnamefont {Hung}},\ and\
  \bibinfo {author} {\bibfnamefont {C.}~\bibnamefont {Chin}},\ }\bibfield
  {title} {\bibinfo {title} {{In situ observation of incompressible
  Mott-insulating domains in ultracold atomic gases}},\ }\href
  {https://doi.org/10.1038/nature08244} {\bibfield  {journal} {\bibinfo
  {journal} {Nature}\ }\textbf {\bibinfo {volume} {460}},\ \bibinfo {pages}
  {995} (\bibinfo {year} {2009})}\BibitemShut {NoStop}%
\bibitem [{\citenamefont {Ho}\ and\ \citenamefont {Zhou}(2010)}]{Ho2010}%
  \BibitemOpen
  \bibfield  {author} {\bibinfo {author} {\bibfnamefont {T.-L.}\ \bibnamefont
  {Ho}}\ and\ \bibinfo {author} {\bibfnamefont {Q.}~\bibnamefont {Zhou}},\
  }\bibfield  {title} {\bibinfo {title} {Obtaining the phase diagram and
  thermodynamic quantities of bulk systems from the densities of trapped
  gases},\ }\href {https://doi.org/10.1038/nphys1477} {\bibfield  {journal}
  {\bibinfo  {journal} {Nature Physics}\ }\textbf {\bibinfo {volume} {6}},\
  \bibinfo {pages} {131} (\bibinfo {year} {2010})}\BibitemShut {NoStop}%
\bibitem [{\citenamefont {Busley}\ \emph {et~al.}(2022)\citenamefont {Busley},
  \citenamefont {Miranda}, \citenamefont {Redmann}, \citenamefont {Kurtscheid},
  \citenamefont {Umesh}, \citenamefont {Vewinger}, \citenamefont {Weitz},\ and\
  \citenamefont {Schmitt}}]{Busley2022}%
  \BibitemOpen
  \bibfield  {author} {\bibinfo {author} {\bibfnamefont {E.}~\bibnamefont
  {Busley}}, \bibinfo {author} {\bibfnamefont {L.}~\bibnamefont {Miranda}},
  \bibinfo {author} {\bibfnamefont {A.}~\bibnamefont {Redmann}}, \bibinfo
  {author} {\bibfnamefont {C.}~\bibnamefont {Kurtscheid}}, \bibinfo {author}
  {\bibfnamefont {K.~K.}\ \bibnamefont {Umesh}}, \bibinfo {author}
  {\bibfnamefont {F.}~\bibnamefont {Vewinger}}, \bibinfo {author}
  {\bibfnamefont {M.}~\bibnamefont {Weitz}},\ and\ \bibinfo {author}
  {\bibfnamefont {J.}~\bibnamefont {Schmitt}},\ }\bibfield  {title} {\bibinfo
  {title} {Compressibility and the equation of state of an optical quantum gas
  in a box},\ }\href {https://doi.org/10.1126/science.abm2543} {\bibfield
  {journal} {\bibinfo  {journal} {Science}\ }\textbf {\bibinfo {volume}
  {375}},\ \bibinfo {pages} {1403} (\bibinfo {year} {2022})}\BibitemShut
  {NoStop}%
\bibitem [{\citenamefont {Carrasquilla}\ \emph {et~al.}(2011)\citenamefont
  {Carrasquilla}, \citenamefont {Becca},\ and\ \citenamefont
  {Fabrizio}}]{Carrasquilla2011}%
  \BibitemOpen
  \bibfield  {author} {\bibinfo {author} {\bibfnamefont {J.}~\bibnamefont
  {Carrasquilla}}, \bibinfo {author} {\bibfnamefont {F.}~\bibnamefont
  {Becca}},\ and\ \bibinfo {author} {\bibfnamefont {M.}~\bibnamefont
  {Fabrizio}},\ }\bibfield  {title} {\bibinfo {title} {Bose-glass, superfluid,
  and rung-mott phases of hard-core bosons in disordered two-leg ladders},\
  }\href {https://doi.org/10.1103/PhysRevB.83.245101} {\bibfield  {journal}
  {\bibinfo  {journal} {Phys. Rev. B}\ }\textbf {\bibinfo {volume} {83}},\
  \bibinfo {pages} {245101} (\bibinfo {year} {2011})}\BibitemShut {NoStop}%
\bibitem [{\citenamefont {Cr\'epin}\ \emph {et~al.}(2011)\citenamefont
  {Cr\'epin}, \citenamefont {Laflorencie}, \citenamefont {Roux},\ and\
  \citenamefont {Simon}}]{Crepin2011}%
  \BibitemOpen
  \bibfield  {author} {\bibinfo {author} {\bibfnamefont {F.~m.~c.}\
  \bibnamefont {Cr\'epin}}, \bibinfo {author} {\bibfnamefont {N.}~\bibnamefont
  {Laflorencie}}, \bibinfo {author} {\bibfnamefont {G.}~\bibnamefont {Roux}},\
  and\ \bibinfo {author} {\bibfnamefont {P.}~\bibnamefont {Simon}},\ }\bibfield
   {title} {\bibinfo {title} {Phase diagram of hard-core bosons on clean and
  disordered two-leg ladders: Mott insulator–luttinger liquid–bose glass},\
  }\href {https://doi.org/10.1103/PhysRevB.84.054517} {\bibfield  {journal}
  {\bibinfo  {journal} {Phys. Rev. B}\ }\textbf {\bibinfo {volume} {84}},\
  \bibinfo {pages} {054517} (\bibinfo {year} {2011})}\BibitemShut {NoStop}%
\bibitem [{\citenamefont {Rosi}\ \emph {et~al.}(2018)\citenamefont {Rosi},
  \citenamefont {Burchianti}, \citenamefont {Conclave}, \citenamefont
  {S.~Naik}, \citenamefont {Roati}, \citenamefont {Fort},\ and\ \citenamefont
  {Minardi}}]{greymolasses}%
  \BibitemOpen
  \bibfield  {author} {\bibinfo {author} {\bibfnamefont {S.}~\bibnamefont
  {Rosi}}, \bibinfo {author} {\bibfnamefont {A.}~\bibnamefont {Burchianti}},
  \bibinfo {author} {\bibfnamefont {S.}~\bibnamefont {Conclave}}, \bibinfo
  {author} {\bibfnamefont {D.}~\bibnamefont {S.~Naik}}, \bibinfo {author}
  {\bibfnamefont {G.}~\bibnamefont {Roati}}, \bibinfo {author} {\bibfnamefont
  {C.}~\bibnamefont {Fort}},\ and\ \bibinfo {author} {\bibfnamefont
  {F.}~\bibnamefont {Minardi}},\ }\bibfield  {title} {\bibinfo {title}
  {{$\Lambda$-enhanced grey molasses on the D2 transition of Rubidium-87
  atoms}},\ }\bibfield  {journal} {\bibinfo  {journal} {Scientific Reports}\
  }\textbf {\bibinfo {volume} {8}},\ \href
  {https://doi.org/DOI:10.1038/s41598-018-19814-z}
  {DOI:10.1038/s41598-018-19814-z} (\bibinfo {year} {2018})\BibitemShut
  {NoStop}%
\bibitem [{\citenamefont {La~Rooij}\ \emph {et~al.}(2022)\citenamefont
  {La~Rooij}, \citenamefont {Ulm}, \citenamefont {Haller},\ and\ \citenamefont
  {Kuhr}}]{larooij2022}%
  \BibitemOpen
  \bibfield  {author} {\bibinfo {author} {\bibfnamefont {A.}~\bibnamefont
  {La~Rooij}}, \bibinfo {author} {\bibfnamefont {C.}~\bibnamefont {Ulm}},
  \bibinfo {author} {\bibfnamefont {E.}~\bibnamefont {Haller}},\ and\ \bibinfo
  {author} {\bibfnamefont {S.}~\bibnamefont {Kuhr}},\ }\bibfield  {title}
  {\bibinfo {title} {A comparative study of deconvolution techniques for
  quantum-gas microscope images},\ }\bibfield  {journal} {\bibinfo  {journal}
  {arXiv:2207.08663}\ }\href {https://doi.org/10.48550/arXiv.2207.08663}
  {10.48550/arXiv.2207.08663} (\bibinfo {year} {2022})\BibitemShut {NoStop}%
\bibitem [{\citenamefont {Weitenberg}\ \emph {et~al.}(2011)\citenamefont
  {Weitenberg}, \citenamefont {Endres}, \citenamefont {Sherson}, \citenamefont
  {Cheneau}, \citenamefont {Schau{\ss}}, \citenamefont {Fukuhara},
  \citenamefont {Bloch},\ and\ \citenamefont {Kuhr}}]{Weitenberg2011}%
  \BibitemOpen
  \bibfield  {author} {\bibinfo {author} {\bibfnamefont {C.}~\bibnamefont
  {Weitenberg}}, \bibinfo {author} {\bibfnamefont {M.}~\bibnamefont {Endres}},
  \bibinfo {author} {\bibfnamefont {J.~F.}\ \bibnamefont {Sherson}}, \bibinfo
  {author} {\bibfnamefont {M.}~\bibnamefont {Cheneau}}, \bibinfo {author}
  {\bibfnamefont {P.}~\bibnamefont {Schau{\ss}}}, \bibinfo {author}
  {\bibfnamefont {T.}~\bibnamefont {Fukuhara}}, \bibinfo {author}
  {\bibfnamefont {I.}~\bibnamefont {Bloch}},\ and\ \bibinfo {author}
  {\bibfnamefont {S.}~\bibnamefont {Kuhr}},\ }\bibfield  {title} {\bibinfo
  {title} {{Single-spin addressing in an atomic Mott insulator}},\ }\href
  {https://doi.org/10.1038/nature09827} {\bibfield  {journal} {\bibinfo
  {journal} {Nature}\ }\textbf {\bibinfo {volume} {471}},\ \bibinfo {pages}
  {319} (\bibinfo {year} {2011})}\BibitemShut {NoStop}%
\bibitem [{\citenamefont {Simon}\ \emph {et~al.}(2011)\citenamefont {Simon},
  \citenamefont {Bakr}, \citenamefont {Ma}, \citenamefont {Tai}, \citenamefont
  {Preiss},\ and\ \citenamefont {Greiner}}]{Simon2011}%
  \BibitemOpen
  \bibfield  {author} {\bibinfo {author} {\bibfnamefont {J.}~\bibnamefont
  {Simon}}, \bibinfo {author} {\bibfnamefont {W.~S.}\ \bibnamefont {Bakr}},
  \bibinfo {author} {\bibfnamefont {R.}~\bibnamefont {Ma}}, \bibinfo {author}
  {\bibfnamefont {M.~E.}\ \bibnamefont {Tai}}, \bibinfo {author} {\bibfnamefont
  {P.~M.}\ \bibnamefont {Preiss}},\ and\ \bibinfo {author} {\bibfnamefont
  {M.}~\bibnamefont {Greiner}},\ }\bibfield  {title} {\bibinfo {title} {Quantum
  simulation of antiferromagnetic spin chains in an optical lattice},\ }\href
  {https://doi.org/10.1038/nature09994} {\bibfield  {journal} {\bibinfo
  {journal} {Nature}\ }\textbf {\bibinfo {volume} {472}},\ \bibinfo {pages}
  {307} (\bibinfo {year} {2011})}\BibitemShut {NoStop}%
\bibitem [{\citenamefont {Ma}\ \emph {et~al.}(2011)\citenamefont {Ma},
  \citenamefont {Tai}, \citenamefont {Preiss}, \citenamefont {Bakr},
  \citenamefont {Simon},\ and\ \citenamefont {Greiner}}]{Ma2011}%
  \BibitemOpen
  \bibfield  {author} {\bibinfo {author} {\bibfnamefont {R.}~\bibnamefont
  {Ma}}, \bibinfo {author} {\bibfnamefont {M.~E.}\ \bibnamefont {Tai}},
  \bibinfo {author} {\bibfnamefont {P.~M.}\ \bibnamefont {Preiss}}, \bibinfo
  {author} {\bibfnamefont {W.~S.}\ \bibnamefont {Bakr}}, \bibinfo {author}
  {\bibfnamefont {J.}~\bibnamefont {Simon}},\ and\ \bibinfo {author}
  {\bibfnamefont {M.}~\bibnamefont {Greiner}},\ }\bibfield  {title} {\bibinfo
  {title} {Photon-assisted tunneling in a biased strongly correlated bose
  gas},\ }\href {https://doi.org/10.1103/PhysRevLett.107.095301} {\bibfield
  {journal} {\bibinfo  {journal} {Phys. Rev. Lett.}\ }\textbf {\bibinfo
  {volume} {107}},\ \bibinfo {pages} {095301} (\bibinfo {year}
  {2011})}\BibitemShut {NoStop}%
\end{thebibliography}%

\newpage


\section*{Supplementary Information}
\label{sec:SI}

\subsection*{State detection probabilities}

\subsection*{Data analysis and post-selection}
\label{sec:Methods-PS)}

We post-select our data sets using two criteria. Firstly, we  exclude 1D systems in which  we observe an atom on the inner site of the potential wall. For  the data presented in Fig.~\ref{fig:JovU}, this occurs in less than $5\%$ of systems. When we apply the bias potential (datasets shown  in Fig.~\ref{fig:Gradient}), this rises to around $20\%$  at the maximum potential gradient. Second, we post-select based on the parity of the atom number of each 1D system.
When we compress the systems by one site, the parity we observe should be the opposite compared to the initially prepared system  since we expect that one site becomes doubly occupied and detected as an empty one.
\begin{figure}[t!]
\centering
\includegraphics[width=\linewidth]{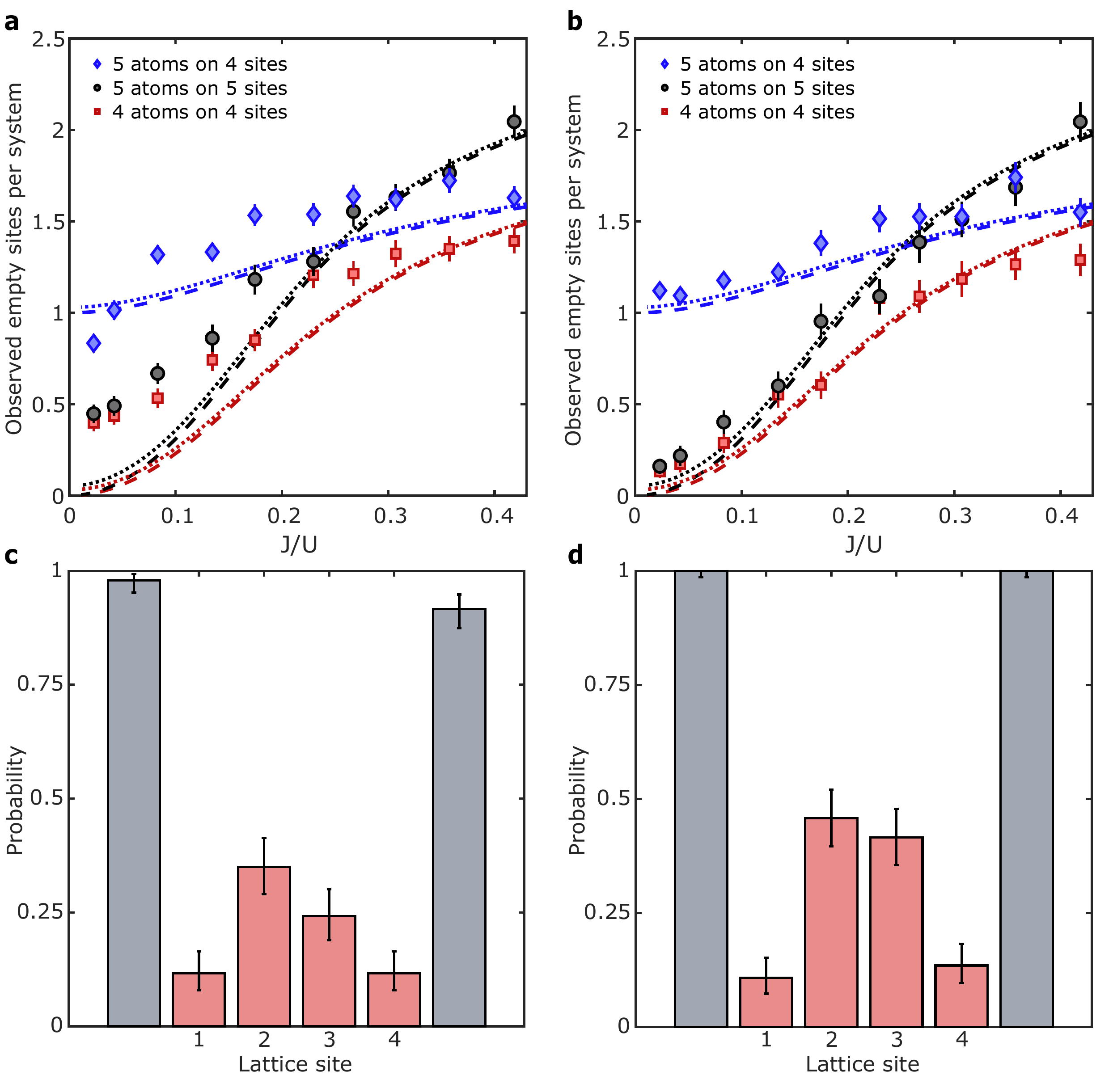}
\caption{\label{fig:postselection}\textbf{Effect of post-selection. a}, Datasets of Fig.~\ref{fig:JovU}a, but without post-selection, \textbf{b} same datasets with post-selection for comparison, \textbf{c},  hole probability for 5 atoms on 4 sites  without post-selection (datasets from Fig.~\ref{fig:DopingSequence}i), \textbf{d} same datasets after post-selection.}
\end{figure}

The fidelity of our system preparation (after Step 1 of the experimental procedure described in Fig.\,\ref{fig:DopingSequence}a) is such that we observe zero empty sites in typically $75\%$ of cases, an incorrect parity in $20\%$ of cases and two empty sites in 5\% of cases. After the whole  experimental procedure (Step 4  in Fig.\,\ref{fig:DopingSequence}a), we find that the observed parity is wrong in 25\%-40\% of the cases, which we attributed to loss of two atoms and particle-hole pair excitations due to heating from intensity noise of the trapping lasers, especially during the intensity ramps. As the parity is conserved a two-atom loss is not detected in the post-selection. Assuming that the probability to lose an atom in our 1D systems is 0.3, then the probability of losing two atoms is 0.1.
We also find that the incommensurate systems require more post-selection indicating that the additional atom is more susceptible to loss which again we attribute to the fact that these systems have a non-gapped excitation spectrum and are more susceptible to  technical noise. Overall, we retain approximately $70\%$ of systems for the data sets shown in Fig.~\ref{fig:JovU}. For the experiments presented in Fig.~\ref{fig:DensityScan}, when we compress the system by more than one site, we retain on average $60\%$ of the datasets. When applying the bias potential in Fig.~\ref{fig:Gradient}, we find that about $55\%$ of the incommensurate systems have the correct parity, and $60\%-75\%$ for the commensurate systems.

We illustrate  the effect of the post-selection process for selected datasets (Fig. \ref{fig:postselection}). For example, in case of the measurement presented in Fig.~\ref{fig:JovU}a in the main text, the post-selection process, makes the datasets match the  $T=0$ theory line much better. However a small offset remains due to the fact that we do not post-select systems where the observed atom number is two less  than what we would  expect. The fraction of the systems with one thermal excitation is typically on the order of $5\%$.

For all datasets presented in this paper, we have used the eight central 1D systems between the barriers because of their
slightly lower entropy.

\subsection*{Details of potential shape}
\label{sec:barrier-shape}
\begin{figure}[!h]
\centering
\includegraphics[width=0.9\columnwidth]{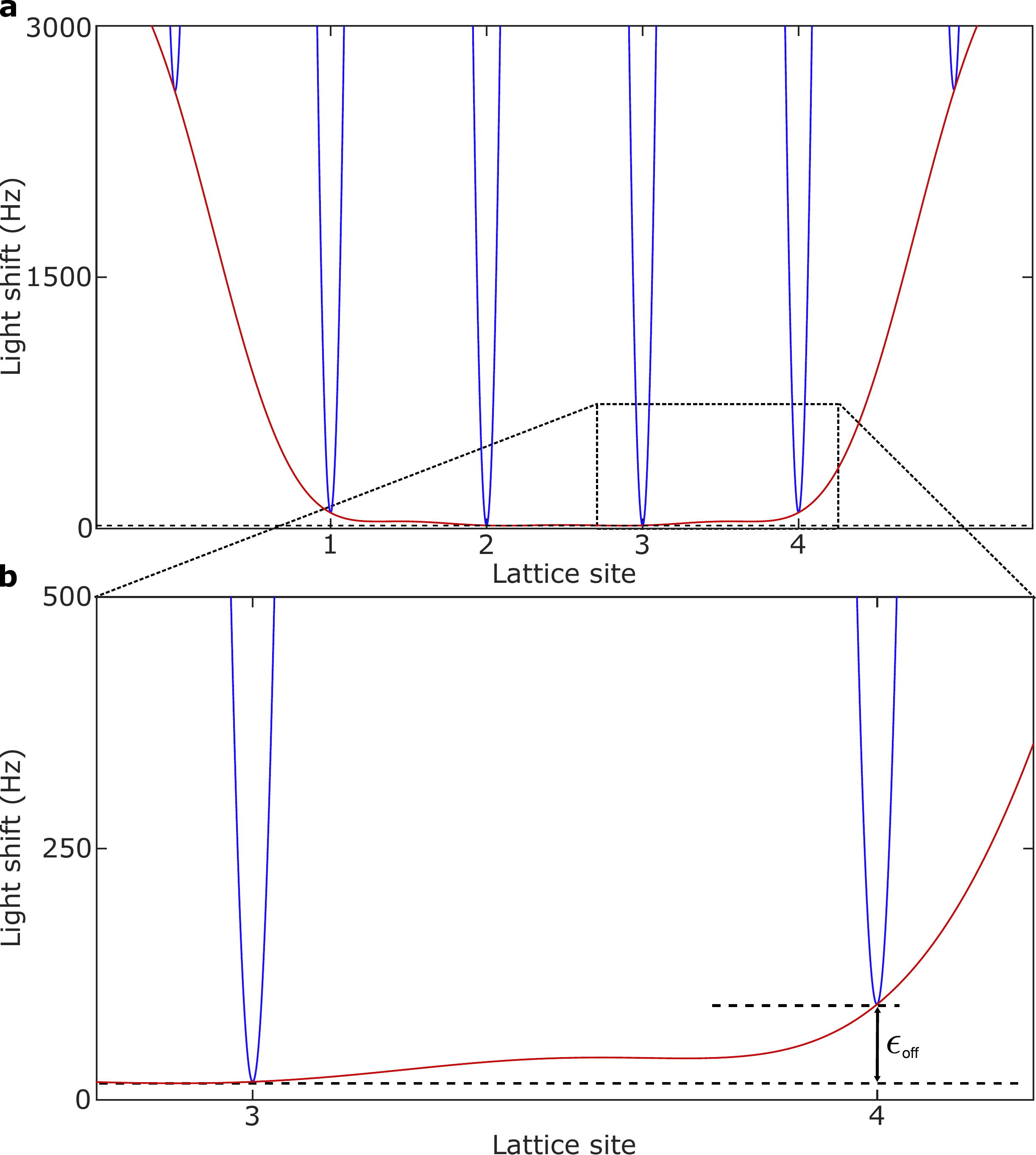}
\caption{\label{fig:potential}\textbf{Effect of the point spread function on the potential shape. a}, Lattice potential of the projected repulsive potential barriers (red) and that of the combined potential including the optical lattice (blue). \textbf{b}, Magnified view of the outer two lattice sites to show the energy offset $\epsilon_{\text{off}}$. }
\end{figure}

In our numerical simulations we take into account the point spread function of the imaging system, which is calculated assuming it is diffraction limited, causing a broadening of the repulsive potential barriers and an energy offset, $\epsilon_{\rm off}$, on the lattice site closest to the barrier (Fig.\,\ref{fig:potential}). For a repulsive barrier producing a maximum light shift of $\Delta_{LS}/h = 3.3(5)\,\mbox{kHz} = 3.5(5) U$,
the energy offset is $\epsilon_{\rm off}=0.027(4) \Delta_{LS}= 2\pi\hbar \times 90(10)$\,Hz. This offset becomes  significant when we apply the bias potential (measurements presented in Fig.\,\ref{fig:Gradient}) as this energy offset opposes the centre of mass shift. For the lattice depth $V_x=16(1)\,E_{\rm r}$ used in Fig.\,\ref{fig:Gradient}, the energy offset on the outermost site of the system is $\epsilon_{\rm off} \approx 0.1\,U \approx 10 J$.

\subsection*{Strong-interactions limit}
The standard Bose-Hubbard model can be solved in the limit of strong interactions via perturbative approaches, instead of the  numerical model we used in this work to simulate the system dynamics. For fixed atom and lattice site numbers, as studied in this work, we can use the limit of $U/J\rightarrow \infty$ (deep lattice) to further restrict the finite Hilbert space. From this, analytical solutions for the low-energy states can be found. For example, in the case of 5 atoms on 3 sites, as we take $U/J\rightarrow \infty$ it is natural to assume that the possibility of three atoms occupying a single site is vanishingly small due to its high onsite-interaction energy of $6U$.
As we have fixed atom and site number, states with two atoms in a single site must be allowed, giving the restricted Hilbert space for the low energy states of $\{\ket{221},\ket{212},\ket{122} \}$. The Hamiltonian is then the kinetic energy term only, with the interaction term being a constant diagonal offset for each basis state. The ground state is then given by
\begin{equation}
    \ket{\psi}_{\textrm{GS}}^{\textrm{5on3}} = \frac{1}{\sqrt{2}} \ket{212} + \frac{1}{2} \left( \ket{221} + \ket{122} \right),
\end{equation}
with the favouring of the atoms being located on the edge of the system.

A similar process can be repeated for each configuration considered in the main text. For example in the case of 6 atoms on 4 sites, we obtain:
\begin{equation}
\begin{aligned}
    \ket{\psi}_\textrm{GS}^\textrm{6on4} = & \frac{1}{2} \left( \ket{2121} + \ket{1212} \right) + \frac{1}{\sqrt{5}} \left( \ket{2112} + \ket{1221} \right) \\ & + \frac{1}{2\sqrt{5}} \left( \ket{2211} + \ket{1122}
 \right),
 \end{aligned}
 \end{equation}
 and the case of five atoms on four sites
 \begin{equation}
     \ket{\psi}_\textrm{GS}^\textrm{5on4} = \mathcal{N} \left[ \ket{1211} + \ket{1121} + \frac{2}{1+\sqrt{5}} \left( \ket{2111} + \ket{1112} \right) \right],
 \end{equation}
with $\mathcal{N} = (1+\sqrt{5})/\left(2\sqrt{2 + \frac{1}{2}\left( 1 + \sqrt{5} \right)^2}\right)$.

\end{document}